\documentclass[aps,prx,twocolumn,groupedaddress,nobibnotes]{revtex4-2}

\usepackage{graphicx}
\usepackage{caption}
\DeclareCaptionLabelSeparator{dot}{. }
\makeatletter
\def\justified{
	\let\\\@normalcr
	\@rightskip\z@skip \rightskip\@rightskip
	\leftskip\z@skip
	\parindent 0em\relax
	\setlength{\parfillskip}{0pt plus 1fil}}
\DeclareCaptionJustification{justified}{\justified}
\usepackage{subcaption}
\usepackage{amsmath}
\usepackage{braket}
\usepackage{units}
\usepackage{ragged2e}
\usepackage{lipsum}
\usepackage[colorlinks,urlcolor=blue ,citecolor=blue ,linkcolor=blue ]{hyperref}
\usepackage{xcolor}
\usepackage{textcomp}
\usepackage{bm}
\usepackage{dsfont}
\usepackage{physics}
\usepackage{titlesec}
\usepackage{float}
\titlespacing*{\section}{0pt}{3.5ex plus 2ex minus .2ex}{2.3ex plus .2ex}
\usepackage{multirow}
\usepackage{placeins}

\usepackage{array}
\newcolumntype{x}[1]{%
>{\centering\hspace{0pt}}p{#1}}%
\newcommand{\tn}{\tabularnewline}

\begin{document}
	
	\title{Robust quantum-network memory \\ based on spin qubits in isotopically engineered diamond}

	\author{C. E. Bradley$^{1,2}$}
	\author{S. W. de Bone$^{1,3}$}
	\author{P. F. W. M\"{o}ller$^{1}$}
	\author{S. Baier$^{1,2}$}
	\author{M. J. Degen$^{1,2}$}
	\author{S. J. H. Loenen$^{1,2}$}
	\author{\\ H. P. Bartling$^{1,2}$}
	\author{M. Markham$^{4}$}
	\author{D. J. Twitchen$^{4}$}
	\author{R. Hanson$^{1,2}$}
	\author{D. Elkouss$^{1}$}
	\author{T. H. Taminiau$^{1,2}$}
	\email{t.h.taminiau@tudelft.nl}
	
	\affiliation{$^{1}$QuTech, Delft University of Technology, 2628 CJ Delft, The Netherlands}
    \affiliation{$^{2}$Kavli Institute of Nanoscience, Delft University of Technology, 2628 CJ Delft, The Netherlands}
    \affiliation{$^{3}$QuSoft, Centrum Wiskunde \& Informatica, 1098 XG, Amsterdam, The Netherlands}
	\affiliation{$^{4}$Element Six Innovation, Fermi Avenue, Harwell Oxford, Didcot, Oxfordshire, OX110QR, United Kingdom}	

	\date{\today}

    \begin{abstract}
	Quantum networks can enable long-range quantum communication and modular quantum computation. A powerful approach is to use multi-qubit network nodes which provide the quantum memory and computational power to perform entanglement distillation, quantum error correction, and information processing. Nuclear spins associated with optically-active defects in diamond are promising qubits for this role. However, their dephasing during entanglement distribution across the optical network hinders scaling to larger systems. In this work, we show that a single $^{13}$C spin in isotopically engineered diamond offers a long-lived quantum memory that is robust to the optical link operation of an NV centre. The memory lifetime is improved by two orders-of-magnitude upon the state-of-the-art, and exceeds the best reported times for remote entanglement generation. We identify ionisation of the NV centre as a newly limiting decoherence mechanism. As a first step towards overcoming this limitation, we demonstrate that the nuclear spin state can be retrieved with high fidelity after a complete cycle of ionisation and recapture. Finally, we use numerical simulations to show that the combination of this improved memory lifetime with previously demonstrated entanglement links and gate operations can enable key primitives for quantum networks, such as deterministic non-local two-qubit logic operations and GHZ state creation across four network nodes. Our results pave the way for test-bed quantum networks capable of investigating complex algorithms and error correction.
	\end{abstract}

	\maketitle

\section{Introduction}

\begin{figure*}
		\includegraphics[width=0.85\linewidth]{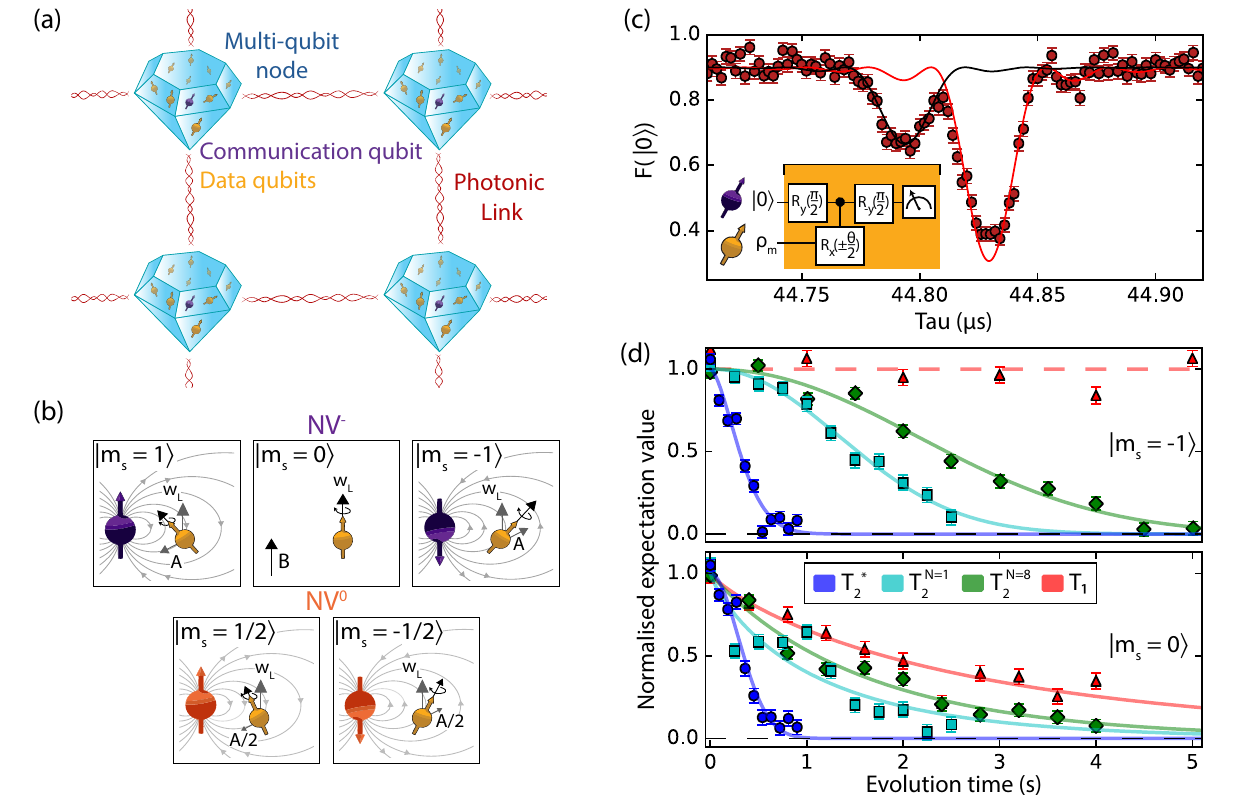}%
		\caption{\label{fig:network} (a) Architecture for distributed quantum information processing using spins in diamond. Multi-qubit nodes are formed from electron spin `communication' qubits (purple) which magnetically couple to nuclear spin `data' qubits (yellow), allowing universal control. The coherent spin-photon interface of the communication qubits enables the heralded creation of remote photonic entanglement between nodes. The data qubits enable long-lived storage and processing of quantum information. The network portrayed here forms a unit cell for distributed quantum error correction based on the surface code \cite{nickerson2013topological,nickerson2014freely,de2020protocols}. (b) Electron-nuclear hyperfine coupling:  nuclear-spin precession at the Larmor frequency around an external magnetic field $\vec{\mathrm{B}}$ is perturbed by the presence of the electron spin. Depending on the NV charge and spin states: NV$^{-}$ (purple), $m_{s} = \{-1,0,+1\}$, or NV$^{0}$ (orange), $m_{s} = \{-1/2,+1/2\}$, the precession frequency and axis are modified. (c) Dynamical decoupling spectroscopy of the NV centre \cite{taminiau2012detection}. The resonance at 44.832 $\mu$s (red) reveals a single $^{13}$C spin with hyperfine parameters $A_\parallel$ = 2$\pi \cdot$ 80(1) Hz and $A_\perp$ = 2$\pi \cdot$ 271(4) Hz. The resonance at 44.794 $\mu$s (black) corresponds to the $^{13}$C spin bath. Solid lines correspond to a theoretical model \cite{supp}. (d) Intrinsic decoherence timescales of the nuclear spin for different electron states. Solid lines are fits (see methods). Dashed lines is a guide to the eye. $N$ denotes the number of spin-echo pulses. All error bars are one standard error.}
\end{figure*}

Quantum networks have the potential to enable a wealth of applications that go beyond classical technologies, including secure communication, quantum sensor networks, and distributed quantum computation \cite{wehner2018quantum,jiang2007distributed,nickerson2013topological,nickerson2014freely}. Such a network might consist of nodes with stationary `communication' qubits that are connected together by entanglement through photonic channels (Fig. \ref{fig:network}(a)). Each node ideally contains multiple additional `data' qubits that can be used to store and process quantum states. Universal operations over the network can then be performed by repeatedly distributing entangled states and subsequently consuming them \cite{jiang2007distributed,nickerson2013topological}.

Large-scale networks and universal quantum computation become possible if imperfections can be overcome through entanglement distillation and quantum error correction \cite{nickerson2013topological,nickerson2014freely}. Besides high-fidelity operations, this requires the faithful storage of quantum states while new entangled states are repeatedly distributed over the network \cite{jiang2007distributed,nickerson2013topological, nickerson2014freely,monroe2014large}. This capability is captured by the (active) `link efficiency', $\eta^{*}_{\mathrm{link}}=r_{\mathrm{ent}}/r_{\mathrm{dec}}$, given by the ratio of the inter-node entanglement generation rate $r_{\mathrm{ent}}$ and the decoherence rate $r_{\mathrm{dec}}$ of the data qubits during network operation \footnote{In prior work, a passive version of the link efficiency, $\eta_{\mathrm{\,link}}$, was evaluated considering $r_{\mathrm{dec}}$ to be the decoherence rate of a two-qubit entangled state during \emph{idling} of the communication link, rather than the single-qubit rate during remote entanglement operation \cite{humphreys2018deterministic}. This metric relates specifically to deterministic entanglement delivery on a clock cycle \cite{humphreys2018deterministic} and bounds $\eta^{*}_{\mathrm{\,link}}$. A large $\eta_{\mathrm{\,link}}$ is a necessary, but not sufficient condition for large quantum networks. While previous experiments reported $\eta_{\mathrm{\,link}}$ in the range of 5-8 \cite{hucul2015modular,humphreys2018deterministic}, the $\eta^{*}_{\mathrm{\,link}}$ in those works was much smaller than 1.}. Without error correction, $\eta^{*}_{\mathrm{\,link}}$ sets the available number of cycles of entanglement distribution, and thereby the depth of protocols and computations that can be performed effectively. While quantum error correction can ultimately increase achievable circuit depths, this will require $\eta^{*}_{\mathrm{\,link}} \gg$ 1 \cite{jiang2007distributed,nickerson2013topological,nickerson2014freely,de2020protocols}.

Various systems have demonstrated basic building blocks for optical quantum networks \cite{ritter2012elementary,hucul2015modular,hensen2015experimental,rosenfeld2017event,stockill2017phase,humphreys2018deterministic,krutyanskiy2019light,stephenson2020high,yu2020entanglement,daiss2021quantum,schupp2021interface}. The nitrogen-vacancy (NV) centre in diamond is a promising platform because it combines a spin-photon interface for heralded remote entanglement \cite{robledo2011high,bernien2013heralded,abobeih2018one}, with access to multiple $^{13}$C nuclear-spin data qubits that can store quantum states for long times \cite{maurer2012room,kalb2018dephasing,bradley2019ten,unden2019revealing,bartling2021coherence,randall2021many,abobeih2021fault}. Entanglement generation rates $r_{\mathrm{ent}}$ larger than the idle qubit decoherence rate have been shown, enabling a single cycle of entanglement delivery deterministically on a clock cycle \cite{humphreys2018deterministic}. As a first step towards exploiting additional computational power in the nodes, experiments with up to two qubits per node demonstrated two-cycle network protocols such as entanglement distillation \cite{kalb2017entanglement} and entanglement-swapping in a three-node network \cite{pompili2021realization,hermans2021qubit}. However, in all these experiments $\eta^{*}_{\mathrm{\,link}} <$ 1. The resulting network operation is inherently probabilistic --- if entanglement generation cycles do not succeed early enough, all previously established quantum states stored in the network memory are lost --- hindering the scaling to larger systems operating over many cycles. Realizing large $\eta^{*}_{\mathrm{\,link}}$ will require higher entanglement rates $r_{\mathrm{ent}}$ through efficient spin photon interfaces \cite{sipahigil2016integrated,riedel2017deterministic,nguyen2019quantum,ruf2021resonant} and/or reduced dephasing of the nuclear spin qubits during the network operation ($r_{\mathrm{dec}}$) \cite{kalb2017entanglement,reiserer2016robust}.

In this work, we focus on the latter challenge. We demonstrate that $^{13}$C spin qubits in isotopically engineered diamond provide robust data qubits for quantum networks. We develop control and single-shot readout of an individual $^{13}$C spin that is weakly coupled to a single NV centre. We then show that an arbitrary quantum state can be stored in this data qubit for over 10$^{5}$ repetitions of a remote entanglement sequence. For current entanglement link success rates, this would imply $\eta^{*}_{\mathrm{\,link}} \approx 10$. We show through numerical network simulations that such a high $\eta^{*}_{\mathrm{\,link}}$ value can enable a next generation of network protocols, including deterministic remote two-qubit gates and the distillation of entangled states over a four-node network, which provides a primitive for surface-code quantum error correction (Fig. \ref{fig:network}(a)) \cite{nickerson2013topological,nickerson2014freely,de2020protocols}. Finally, we identify ionisation of the NV centre as a limiting mechanism and show that the data qubit can be protected through fast controlled resetting of the NV charge state. When combined with recent progress with optical cavities towards enhanced entanglement rates \cite{riedel2017deterministic,ruf2021resonant} and high-fidelity quantum gates \cite{casanova2015robust,bradley2019ten,abobeih2021fault}, these results indicate that nuclear spins in isotopically engineered samples provide a promising test-bed for quantum networks. 

\section{Results}

\subsection{System overview}

We consider quantum network nodes consisting of a single NV centre coupled to multiple $^{13}$C nuclear spins (Fig. \ref{fig:network}(a)). The optically active NV electron spin acts as a communication qubit and the $^{13}$C nuclear spins are additional data qubits that also function as quantum memory to store states while new entanglement links between NV centres are established. A key feature of this system is that the $^{13}$C spin dynamics depend on the NV electron spin state (Fig. \ref{fig:network}(b)). This is captured by the Hamiltonian for a single $^{13}$C spin:
\begin{equation}\label{eq:Hgeneral}
H = \omega_{L} I_{z} + A_{\parallel} m_{s} I_{z} + A_{\bot} m_{s} I_{x}.
\end{equation}
Here, we have made the secular approximation. $\omega_L = \gamma_{\mathrm{C}} B_z$ is the nuclear spin Larmor frequency, where $\gamma_{\mathrm{C}}$ is the $^{13}$C gyromagnetic ratio and $B_z$ is the external magnetic field along the NV axis (here $B_z$ = 47 G). $I_{\alpha}$ are the nuclear spin-1/2 operators, while $m_{s}$ is the spin-z projection of the electron spin. In the NV$^{-}$ charge state that is used for network operation, $m_{s}\,\in\,\{-1,0,1\}$, from which we define a qubit in the $\{-1,0\}$ basis ($:=\{\ket{1},\ket{0}\}$). Additionally, however, stochastic ionisation events can convert NV$^{-}$ to the NV$^{0}$ state with $m_{s}$ $\in\,\{\nicefrac{-1}{2},\nicefrac{1}{2}\}$. 

As seen from Eq. \ref{eq:Hgeneral}, the $^{13}$C spin undergoes different precession dependent on the NV charge- and spin-state. This conditional precession enables complete control over the $^{13}$C spins by controlled inversions of the electron NV spin \cite{bradley2019ten}. Uncontrolled electron-spin dynamics, however, induce additional dephasing of the $^{13}$C spin, which sets a limit on the achievable link efficiency $\eta^{*}_{\mathrm{\,link}}$ \cite{jiang2008coherence,blok2015towards,reiserer2016robust,kalb2018dephasing}. Our approach to creating robust memory is to reduce the coupling between the NV electron spin and the $^{13}$C data qubits by reducing the $^{13}$C concentration \cite{balasubramanian2009ultralong,maurer2012room,pfender2017nonvolatile}. While this approach results in slower gate speeds, these are not the rate-limiting step in current network experiments \cite{pompili2021realization}. Note that, in the case of $^{13}$C-spin-bath-limited decoherence, the coherence times increase proportionally with the reduction in coupling strength, such that no intrinsic reduction in gate fidelity is expected \cite{maze2008electron,balasubramanian2009ultralong}.


Our experiments are performed on a single NV centre in a type-IIa isotopically-purified diamond (targeted $^{13}$C concentration of 0.01\%) at a temperature of 4 K. The hardware setup and NV centre properties are described in the methods section.

Dynamical decoupling (DD) spectroscopy with the NV centre \cite{taminiau2012detection} reveals coupling to an isolated $^{13}$C spin, along with the wider spin bath (Fig. \ref{fig:network}(c)). We characterise the electron-nuclear hyperfine components parallel (perpendicular) to the NV axis to be $A_\parallel$ = 2$\pi \cdot$ 80(1) Hz ($A_\perp$ = 2$\pi \cdot$ 271(4) Hz) \cite{taminiau2012detection,boss2016one}. In comparison to previous studies in natural $^{13}$C abundance samples \cite{reiserer2016robust,kalb2018dephasing}, the electron-nuclear coupling strength is reduced by approximately two orders of magnitude, as expected from the isotopic concentration. 

We realise universal control over the electron-$^{13}$C two-qubit system by microwave (MW) and radio-frequency (RF) single-qubit gates, and a DD-based electron-nuclear two-qubit gate \cite{taminiau2014universal,bradley2019ten}. Furthermore, we develop repetitive readout of the nuclear spin to improve the single-shot readout fidelity \cite{supp,maurer2012room,dreau2013single,liu2017single,kalb2016experimental,pfender2019high,cujia2019tracking}, giving a maximum state preparation and measurement (SPAM) fidelity of 91(1)\% \cite{supp}. For the measurements reported here we focus on the system dynamics rather than maximising the fidelities, and thus use a faster initialisation procedure to optimise the signal-to-noise ratio, with lower SPAM fidelity of 79.4(9)\% (see supplement \cite{supp}). The fidelities are predominantly limited by electron-spin decoherence during two-qubit gates, likely arising from the high concentration of P1 centre impurities in this specific device ($\sim$ 75 ppb \cite{supp,degen2021entanglement}). Multi-qubit registers with comparable two-qubit gate fidelities to those achieved in natural-abundance $^{13}$C devices ($\sim$99\% \cite{bradley2019ten}) are expected to be feasible with reduced impurity concentration in future diamond growth.

\subsection{Memory robustness while idling}

We first investigate the intrinsic decoherence processes of the $^{13}$C nuclear spin, i.e. when the electron spin is idle. The electron spin state affects spin-bath dynamics, as flip-flop interactions between nuclear spins are suppressed when the hyperfine interaction is present (the `frozen core' effect) \cite{khutsishvili1962spin,guichard2015decoherence, bradley2019ten}. Therefore, we separately characterise the decoherence timescales for the NV electron in the $\ket{1}$ or $\ket{0}$ states. 

Figure \ref{fig:network}(d) summarises the measured timescales. The observation of vastly longer $T_{1,e=\ket{1}} (\gg 5$ s) than $T_{1,e=\ket{0}} (= 2.8(2)$ s) indicates that the spin relaxes primarily through flip-flop interactions with other $^{13}$C spins. The measured $T_{1,e=\ket{0}}$ is approximately 30-100 times longer than typical values for natural abundance samples, consistent with the expectation of a linear dependence on isotopic concentration \cite{maze2008electron,balasubramanian2009ultralong}. 

The measured dephasing times are $T_{2,e=\ket{1}}^{*}$ = 0.38(1) s and $T_{2,e=\ket{0}}^{*}$ = 0.42(2) s. We first isolate the timescale associated with couplings to the nuclear spin bath by performing a spin-echo experiment on all $^{13}$C spins, decoupling the $^{13}$C bath from other processes but leaving their mutual interactions unperturbed. We find a characteristic timescale, $T_{2,^{13}\mathrm{C-bath}}^{*}$ = 0.66(3) s (see supplement \cite{supp}). The contribution from P1 impurities can similarly be estimated using the NV electron dephasing time during double electron-electron resonance measurements with the P1 impurities, $T_{2,\mathrm{P1-bath}}^{*, e}$ = 230 $\mu$s \cite{degen2021entanglement}. From the respective gyromagnetic ratios, $\gamma_{\mathrm{C}}$, $\gamma_{\mathrm{e}}$, we infer $T_{2,\mathrm{P1-bath}}^{*}$ = 0.61 s. Combining these processes gives an estimate for $T_{2}^{*} \sim$ 0.45 s, close to the measured values. 

We finally consider the application of spin echoes which mitigate quasi-static noise. The single-echo $T_{2,e=\ket{0}}^{N=1}$ (= 1.11(8) s) and eight-echo $T_{2,e=\ket{0}}^{N=8}$ (= 1.62(9) s) are likely limited by the $m_{s}=0$ nuclear $T_{1}$. The observation of some underlying structure in these measurements is likely due to the dynamics of other proximal $^{13}$C spins \cite{abobeih2018one}. In contrast, for the electron $\ket{1}$ state, the nuclear spin frozen core has previously been shown to significantly enhance $^{13}$C spin-echo times \cite{bradley2019ten}. While $T_{2,e=\ket{1}}^{N=1}$ (= 1.82(6) s) and $T_{2,e=\ket{1}}^{N=8}$ (= 2.91(8) s) do exceed $T_{2,e=\ket{0}}^{N=1,8}$ respectively, they are shorter than would be predicted from the $^{13}$C concentration alone ($\sim$ 1 minute). We ascribe this to the presence of the P1 bath, which exhibits faster dynamics which are not well mitigated by echo pulses. 

\subsection{Memory robustness during the remote-entanglement sequence}

\begin{figure}[ht!]
	\includegraphics[width=0.85\linewidth]{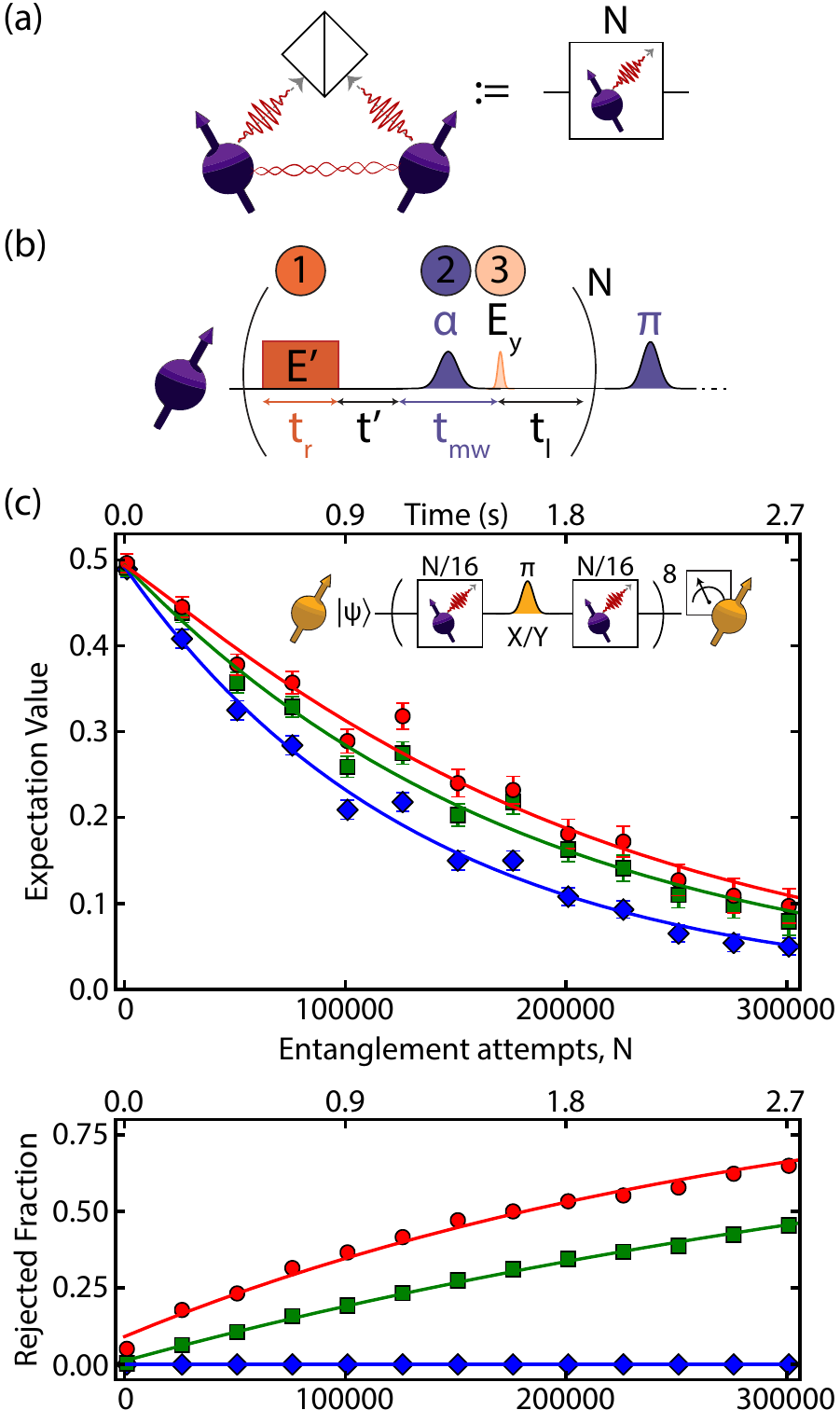}%
	\caption{\label{fig:Qmem} Memory robustness under remote-entanglement generation. (a) Sketch of a remote-entanglement process. After generation of local spin-photon entanglement at each node (see (b)), a detection event at a central beamsplitter heralds remote entanglement. Due to loss processes, the protocol is probabilistic. (b) Remote entanglement primitive: (1) The electron spin is reset by an optical pulse of duration $t_{\mathrm{r}}=5$ $\mu$s. (2) After a time $t'=200$ ns, a microwave pulse of duration $t_{\mathrm{mw}}=2.8$ $\mu$s prepares a spin superposition state. (3) An optical $\pi$-pulse generates spin-photon entanglement, after which a time $t_{\mathrm{l}}=1$ $\mu$s is required to determine whether the attempt succeeded. In case of success, a microwave $\pi$-pulse is used to preserve the electron coherence. E' and E$_{\mathrm{y}}$ denote optical transitions (methods). (c) Data qubit expectation value in the encoded state eigenbasis, as a function of the number of applied primitives (see (b)), averaged over the six cardinal states. Diamonds correspond to no post-selection, squares and circles correspond to post-selection on measuring $\geq$1 and $\geq$5 photons in a CR check performed at the end of the sequence. Solid lines are fits (see methods). Lower panel shows the fraction of rejected data for the three cases.}
\end{figure}

Armed with a characterisation of the intrinsic decoherence processes, we now turn to the remote entanglement sequence shown in Figs. \ref{fig:Qmem}(a,b). This entangling primitive is compatible with single-photon schemes used in recent NV network experiments \cite{cabrillo1999creation,humphreys2018deterministic,pompili2021realization}. In this work, we implement the protocol on a single network node to investigate its effect on the $^{13}$C data qubit.

As discussed in the previous sections, imperfect knowledge of the electron spin state leads to spurious phases acquired by the nuclear spin. Within each entangling attempt, there are a number of processes which can lead to such dephasing: the stochastic nature of the electronic spin reset time, infidelities in the initialised electron spin state, excited state spin-flips after the optical $\pi$-pulse, and MW pulse errors. Reducing the electron-nuclear coupling strength enhances the robustness to such events. 

We implement the entangling protocol experimentally. For simplicity, we omit the optical $\pi$-pulses, but note that the small associated electron spin-flip probability is expected to play a negligible effect on the $^{13}$C decoherence \cite{reiserer2016robust,kalb2018dephasing}. As a trade-off between the optical reset rate and stochastic ionisation to NV$^{0}$ \cite{robledo2010control,siyushev2013optically}, spin reset is performed using 30 nW of optical power for 5 $\mu$s, resulting in an initialisation fidelity $\geq$ 98\% \cite{supp}. For the MW pulse used to create the electron superposition state, we implement a weak multi-tone driving pulse ($\Omega_\mathrm{Rabi}$ $\sim 93$ KHz), see supplement \cite{supp}. This comes at the expense of a time overhead, but mitigates heating. We set the MW rotation angle to $\pi$/2, for which dephasing due to the entangling primitive is expected to be maximal for the given sequence \cite{supp}. Finally, we focus on networks for distributed quantum computation, rather than long-distance communication, and assume that the distance between network nodes is small ($<$ 100 m). Thus, any decision logic can be completed within 1 $\mu$s. With these choices, the total duration of each primitive is 9 $\mu$s.

We prepare the nuclear spin in each of the six cardinal states ($\Psi_{i} \in$ \{$\ket{\mathrm{x}}$,$\ket{\mathrm{-x}}$,$\ket{\mathrm{y}}$,$\ket{\mathrm{-y}}$,$\ket{\uparrow}$,$\ket{\downarrow}$\}), apply N repetitions of the primitive, and measure the expectation value in the associated eigenbasis, $\langle {\hat{\sigma}_{j}} \rangle$ = Tr($\rho \, \hat{\sigma}_{j}$), where $\hat{\sigma}_{j}$ are nuclear spin Pauli operators, j $\in$ \{x,y,z\} (Fig. \ref{fig:Qmem}(c)). We interleave one cycle of XY8 decoupling pulses on the nuclear spin to mitigate dephasing. Without any post-selection of the data we find that the data qubit can preserve an arbitrary quantum state for $N_{\mathrm{1/e}}$ = 1.33(4)$\cdot$10$^{5}$ entangling primitives (1.20(4) s of continuous entanglement attempts).
 
We now consider the effects of ionisation and spectral diffusion on the data. After each experimental run, we perform two-laser probe measurements (E', E$_{\mathrm{y}}$ transitions), denoted charge-resonance (CR) checks (see supplement \cite{supp}) \cite{bernien2013heralded}. The number of photons detected in this check is used to verify that the NV remained in NV$^{-}$ and on resonance throughout the experiment. By varying the post-selection threshold for the CR check, we can reject measurements in which the NV ionised or underwent spectral diffusion with increasing confidence. We find a further improvement in the data qubit lifetime when omitting these cases, showing that they are currently a significant limitation (Fig. \ref{fig:Qmem}(c)). After accounting for ionisation and spectral diffusion, we fit a decay time $N_{\mathrm{1/e}}$ = 2.07(8)$\cdot$10$^{5}$. Considering spin superposition- and eigen-states separately, we find $N_{\mathrm{1/e,\, xy}}$ = 1.90(8)$\cdot$10$^{5}$ [1.71(7) s] and $N_{\mathrm{1/e,\, z}}$ = 2.6(2)$\cdot$10$^{5}$ attempts [2.4(2) s].

In the supplement, we additionally present results for the decay of a superposition state when using strong microwave pulses, so that the sequence is shorter (primitive duration 6.3 $\mu$s) \cite{supp}. We find a further improved corrected decay constant of $N_{\mathrm{1/e, \, x}}$ = 4(1)$\cdot$10$^{5}$ attempts [2.6(6) s]. However, in this case, we cannot rule out that heating due to the strong pulses changes the repumping dynamics, although the observation of similar ionisation statistics suggests that this is not the case. Such heating can be addressed by improved device engineering in future work. 

For both primitive durations (6.3 $\mu$s and 9 $\mu$s), the measured decoherence timescales are comparable to those arising from intrinsic spin-bath dynamics. This suggests that the entanglement sequence only weakly increases the dephasing of the $^{13}$C spin. 
 
We use Monte Carlo simulations to model the nuclear spin dephasing induced by the entanglement attempts, taking into account all known control errors (see supplement \cite{supp}) but neglecting the intrinsic decoherence rates \cite{kalb2018dephasing}. We find that even using pessimistic parameters, an $A_{\parallel}$ = 2$\pi \cdot$ 80 Hz coupled nuclear spin is predicted to retain a fidelity of 77.6(4)\% with respect to an initial superposition state after 10$^{6}$ entangling attempts. This finding is consistent with the interpretation of spin-bath limited decoherence in the present experiment. 

Our simulations also show that the dephasing infidelity after 10$^{6}$ entangling attempts can be reduced below 1\% by using optimised primitives incorporating an additional electron spin echo (see supplement \cite{supp}). Further improvement of the performance can be realised by reducing the entangling primitive duration, such that more attempts can be performed within the intrinsic decoherence timescales. These intrinsic timescales may themselves be extended by optimised decoupling schemes \cite{choi2020robust} as well as using samples with lower nitrogen defect concentrations. It is clear, however, that ionisation currently is a limiting factor. We next probe the nuclear spin dynamics after such ionisation events to understand if they can be mitigated.

\begin{figure}[ht!]
	\includegraphics[width=0.85\linewidth]{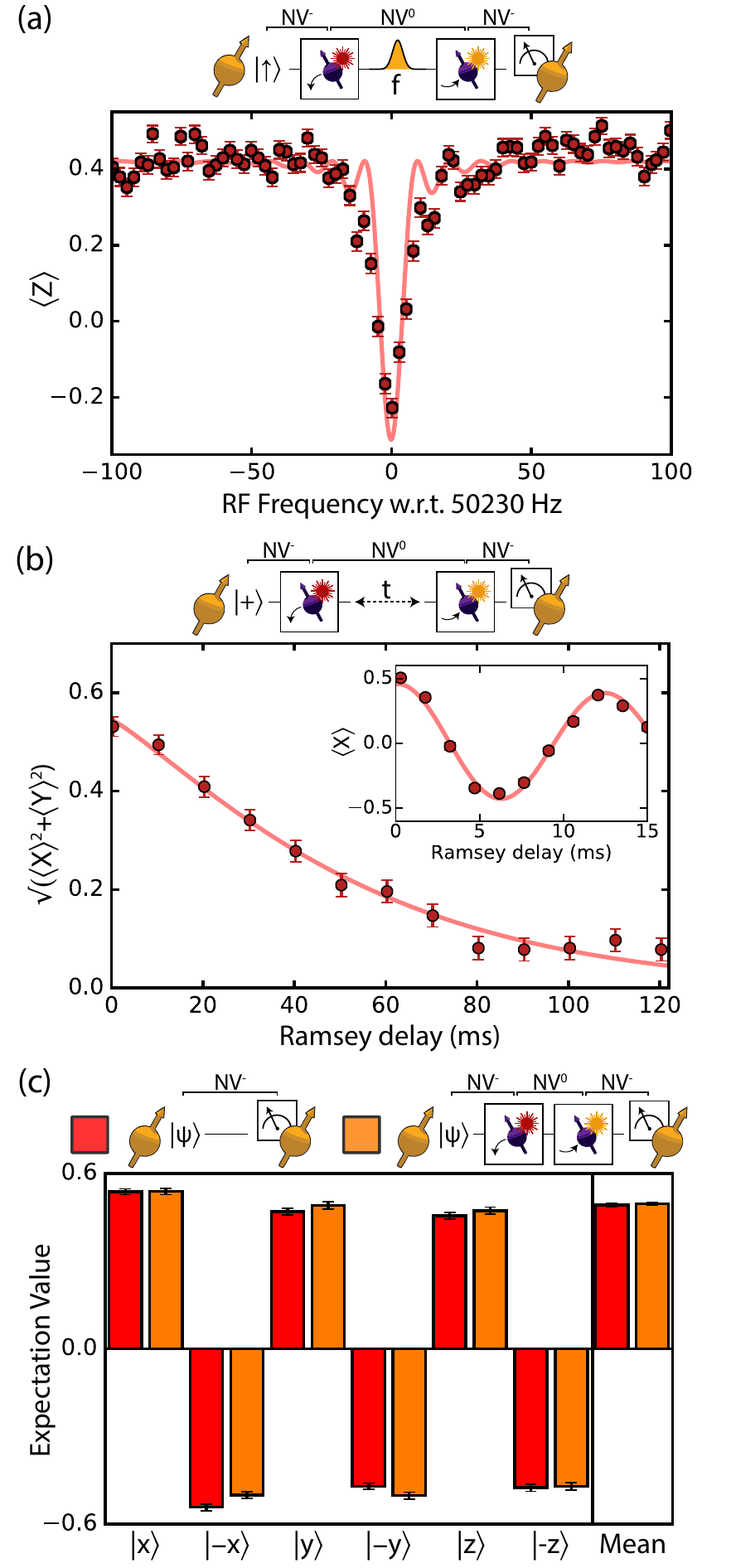}%
	\caption{\label{fig:NV0C13} (a) Nuclear magnetic resonance spectroscopy after heralded preparation of NV$^{0}$. Solid line is a fit (see methods). (b) Free induction decay of the nuclear spin coherence in NV$^{0}$. Inset: Ramsey fringes measured during the first 15 ms of free evolution (artificial detuning of 80 Hz with respect to 50230 Hz). Solid lines are fits (see methods). (c) Protection of an arbitrary data qubit state under NV charge cycling. Orange (red) bars correspond to measurements with a charge-state cycle (idling in NV$^{-}$) between nuclear spin initialisation and readout. The data is post-selected on finding the NV in the negative charge state at the end of the measurement (see text).} 
\end{figure}

\subsection{Mitigating ionisation}

The above section indicates that ionisation of the NV centre limits achievable memory performance. In the above experiments, and in previous work \cite{reiserer2016robust,kalb2018dephasing,pompili2021realization}, an ionisation event (NV$^{-}$ $\rightarrow$ NV$^{0}$) at some point in the remote entanglement sequence causes complete dephasing of the $^{13}$C spins. This is because such events occur stochastically, the subsequent electron spin dynamics are unknown, and, as discussed above (see Eq. \ref{eq:Hgeneral} and Fig. \ref{fig:network}(b)), each electron spin state causes different $^{13}$C spin evolution. Ionisation rates lower than those presented here can be achieved, because single-photon entangling schemes typically use unbalanced spin-superposition states for which the optical excitation probabilities are much smaller \cite{humphreys2018deterministic,pompili2021realization}. However, because weaker optical powers result in longer sequence times and the optical $\pi$-pulses cannot be removed, it is challenging to completely prevent ionisation. A promising approach is to develop techniques which make the data qubit robust to this process. 

In order to study the effect of charge-state switching, we implement resonant optical pulses that induce ionisation (NV$^{-}$ $\rightarrow$ NV$^{0}$, 1 ms) and recharging (NV$^{0}$ $\rightarrow$ NV$^{-}$, 1 ms). Combining these pulses with CR checks, we can use heralding and post-selection to perform verified charge-state switching from NV$^{-}$$\rightarrow$NV$^{0}$$\rightarrow$NV$^{-}$ (see supplement \cite{supp}).

We first combine the verified charge-state switching protocol with nuclear spin control to investigate the properties of the $^{13}$C spin for the NV$^0$ state. After preparing NV$^{-}$ and initialising the nuclear spin in the state $\ket{\uparrow}$, we apply the ionisation pulse, and proceed with the experiment if a subsequent CR check heralds preparation in NV$^0$. We then apply a nuclear spin RF $\pi$-pulse with a Rabi frequency of 5.4(2) Hz, for which we sweep the carrier frequency $f$. After applying the recharge pulse, we read out the nuclear spin in the Z-basis and post-select the data on finding the NV in NV$^-$. As shown in Fig. \ref{fig:NV0C13}(a), we observe a single nuclear spin resonance at a frequency of 50229.8(1) Hz, which matches the $^{13}$C Larmor frequency. The observation of a single transition, rather than two transitions at $f = \omega_{L} \pm A_{\parallel}/2$, suggest a fast averaging over the two NV$^0$ electron spin states, akin to motional narrowing \cite{maurer2012room,pfender2017nonvolatile}. 

To characterise the nuclear spin dephasing, we prepare the nuclear spin in a superposition state, and let it evolve freely while switching to NV$^{0}$ for a variable time. The inset of Fig. \ref{fig:NV0C13}(b) reveals coherent oscillations, corresponding to precession of the nuclear spin at a frequency of 50231(1) Hz. The nuclear spin coherence $(\langle X \rangle ^{2} + \langle Y \rangle ^{2})^{1/2}$ decays as $T_{2,\mathrm{NV0}}^{*} = 57(3)$ ms. This dephasing time is significantly shorter than for NV$^-$ ($T_{2,\mathrm{NV-}}^{*} \sim 0.4$ s, Fig. \ref{fig:network}(d)), indicating that the NV$^0$ state introduces an additional dephasing mechanism. The fitted decay exponent of $n=1.2(1)$ (see methods) matches the value of $n=1$ expected for the motional-narrowing-like regime \cite{maurer2012room}.

\begin{figure*}[ht!]
		\includegraphics[width=1\linewidth]{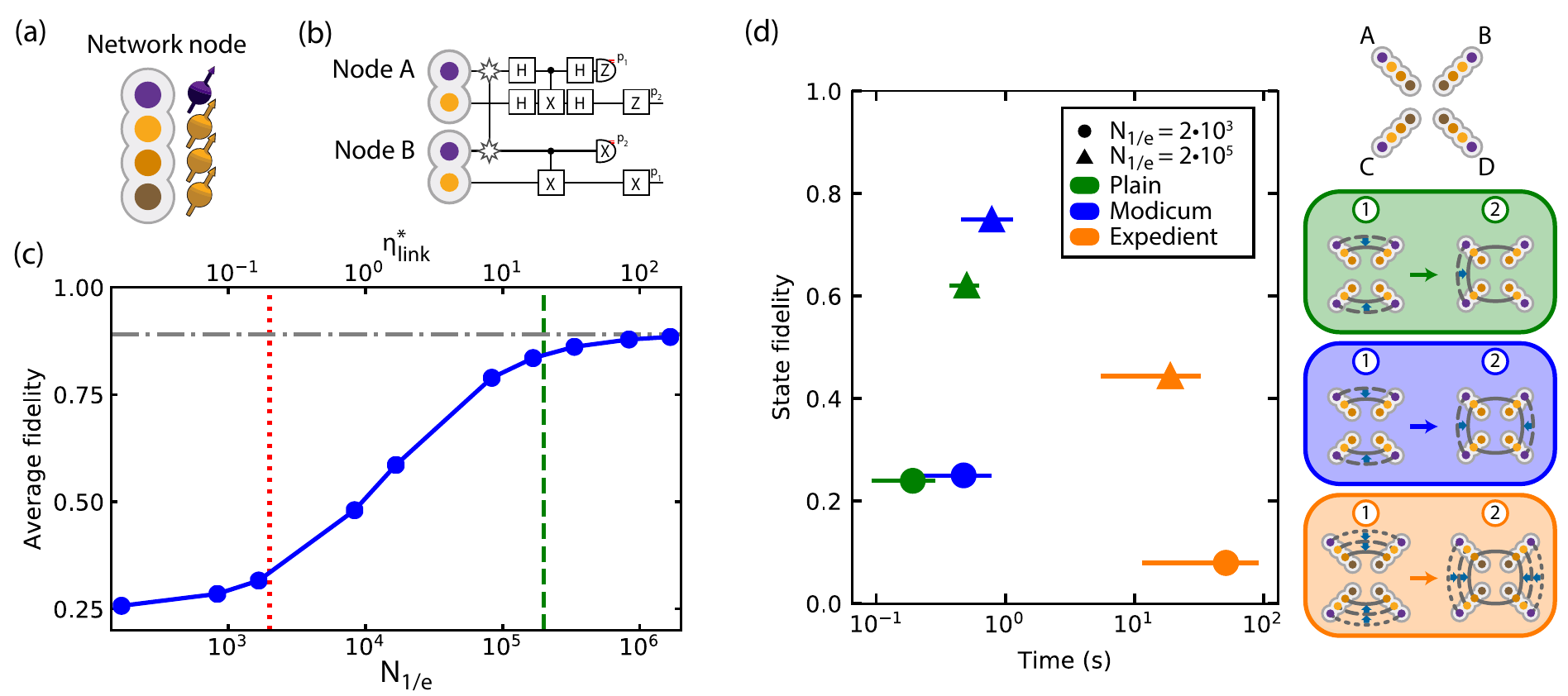}%
		\caption{\label{fig:sims} (a) Diagram of a single network node comprising an electron spin (purple) and up to three nuclear spins (orange tones). (b) Quantum circuit for a CNOT operation between nuclear spins in separate network nodes. $p_{1,2}$ are feed-forward operations. The gate is performed deterministically: the nuclear-spin states are prepared, then entanglement is distributed, and the gate is performed independent of the number of entanglement attempts required and for all measurement outcomes. (c) Simulated performance of the non-local CNOT protocol (see (b)) for a range of data qubit lifetimes. The red and green lines mark memory lifetimes of $N_{\mathrm{1/e}}$ = 2$\cdot$10$^{3}$ and 2$\cdot$10$^{5}$, respectively. The grey line is the limit of the average CNOT fidelity, $F_{\mathrm{av}} = 0.89$, set by error sources other than memory decoherence, see text. Error bars are smaller than markers. (d) Simulated performance for creation of a GHZ state between four network nodes, for three different protocols as schematised on the right. The innermost qubits in the schematics could host the encoded logical information of the distributed surface code, but are not used here. Further details of each protocol are given in the supplement \cite{supp}. The horizontal error bars in (d) indicate the distribution of durations taken for each protocol to succeed (68.2\% confidence intervals).}
\end{figure*}

The observation of a single nuclear spin frequency and a decreased $T_2^*$ are consistent with a rapidly fluctuating NV$^{0}$ electron spin state. In this regime, the dephasing time scales as $T_{2}^{*} = 8/(A_{\parallel}^{2} \cdot T_{1}^{\mathrm{NV0}}$) \cite{maurer2012room,pfender2017nonvolatile}. Thus, we can extract $T_{1}^{\mathrm{NV0}}$ = 570(30) $\mu$s. This value deviates from another recent cryogenic measurement ($T_{1}^{\mathrm{NV0}}=1.51(1)$s) \cite{baier2020orbital}. It is likely that this discrepancy is either due to the much higher magnetic field ($B_{z}=1850$ G) or absence of a P1 bath in that work.

The observed nuclear spin dephasing time in NV$^{0}$ is much longer than the time needed to recharge to NV$^{-}$. We now show that this enables the protection of the data qubit from ionisation events. We prepare the $^{13}$C in each of the six cardinal states, and compare the expectation values obtained when applying the heralded ionisation and recharging process prior to measurement, or when measuring immediately after state preparation (Fig. \ref{fig:NV0C13}(c)). Here, to minimise dephasing during the ionisation and recharging processes, we shorten the ionisation pulse length to 0.3 ms, and use a higher-power 0.5 ms recharging pulse (see supplement \cite{supp}). 

Without post-selecting cases in which the NV returned to the bright state, we find a reduction of the spin expectation values of 8(1)\% (averaged over the cardinal states) when applying the charge-state cycle. Upon post-selection, we find a negligible difference between the cases (the charge-cycle scenario outperforms immediate measurement by 1(1)\%), see Fig. \ref{fig:NV0C13}(c). We thus attribute the majority of the observed infidelity to the finite recharging probability, which is likely limited by spectral diffusion arising from the high impurity concentration in the present sample (see supplement \cite{supp}). We estimate that the NV is in the NV$^{0}$ state for an average of $\sim$ 500 $\mu$s \cite{supp}, for which the $^{13}$C dephasing from the measured $T_{2,\mathrm{NV0}}^{*}$ is $\sim$ 0.5\%, within the uncertainty of the present measurements.

Incorporating such ionisation protection within a network protocol requires a few further considerations. First, the average nuclear spin precession frequency during the entangling primitives should match the precession for NV$^{0}$, else additional dephasing will occur at the difference frequency until recharging occurs. This can be realised by alternating the choice of NV$^{-}$ qubit-basis $m_{s} = \{0,-1\}, \{0,+1\}$ between (or during) entangling attempts. Second, re-charging within the entangling sequence can be realised by adding auxiliary 575 nm excitation during all NV$^{-}$ spin-reset pulses, which prevents extended periods in which the NV resides in the neutral charge-state \cite{humphreys2018deterministic}. A drawback of this approach is that extensive application of yellow light may induce spectral diffusion due to ionisation of the P1 bath or other impurities \cite{heremans2009generation}. A less intrusive approach is to check the charge- and resonance status of the NV centre during the entangling attempts. This could be done passively by monitoring the emitted light either during spin initialisation (in the zero-phonon line, ZPL, and/or the phonon sideband, PSB), or after the optical $\pi$-pulses (in the PSB only, which would not perturb useful quantum information). Alternatively, explicit CR checks could be periodically performed. Efficiently scheduling such checks into a full remote entangling protocol is non-trivial, but the timescales associated with ionisation, data qubit dephasing in NV$^{0}$, CR-checking and recharging suggest that such an approach is feasible.

We note that the NV$^{0}$ spin-$T_{1}$ may increase with reduction of impurities, or under different magnetic fields, which would result in faster nuclear spin dephasing. This challenge might be overcome by either inducing fast NV$^{0}$ spin-flips using microwave driving, or by performing feedback correction using recently demonstrated NV$^{0}$ optical spin-readout at high magnetic fields \cite{baier2020orbital}.

\subsection{Numerical analysis of quantum network protocols}

Finally, we investigate the projected performance of a number of quantum network protocols when combining the improved memory lifetime demonstrated in this work with previously demonstrated entanglement links and gate operations (in other devices). We focus on non-local two-qubit operations and the creation of distributed four-qubit GHZ states, a key building block for the distributed surface-code \cite{nickerson2013topological,nickerson2014freely} (Fig. \ref{fig:network}(a)). 

These investigations are based upon density-matrix simulations of noisy quantum circuits. We use the following set of parameters to model the dominant error sources, which are either the measured state-of-the-art (on a variety of devices) or near-term predictions (denoted by *): two-qubit gates (1\% infidelity \cite{bradley2019ten}), NV optical measurement (1\% infidelity* \cite{hensen2015experimental}), and networked (NV-NV) Bell-pair creation (10\% infidelity* \cite{pompili2021realization}). We account for the finite duration associated with each operation, including probabilistic NV-NV remote entanglement (duration 6 $\mu$s, success probability 0.01\%* \cite{pompili2021realization}), and for decoherence of electron and nuclear spin qubits across all operations \cite{abobeih2018one,bradley2019ten}. We do not take NV ionisation into account. A detailed summary of the models and parameters used can be found in the supplement \cite{supp}, and the code implementation is given in a linked repository \cite{repo}.

In Figs. \ref{fig:sims}(a,b,c), we show the quantum circuit and simulated average fidelity $F_{\mathrm{av}}$ \cite{nielsen2002simple} for a deterministic CNOT operation between nuclear spins in separate network nodes. For $N_{\mathrm{1/e}}$ = 2$\cdot$10$^{3}$ (see e.g. previous work \cite{pompili2021realization}), the projected average fidelity is $F_{\mathrm{av}} \approx$ 0.33, close to that of a completely random operation ($F_{\mathrm{av}}$ = 0.25); the nuclear spin decoheres on a timescale shorter than the average time taken to create an entangled resource state ($\eta^{*}_{\mathrm{\,link}} < 1$). In contrast, for $N_{\mathrm{1/e}}$ = 2$\cdot$10$^{5}$ as obtained here, we find a projected gate fidelity of $F_{\mathrm{av}} \approx$ 0.86, which approaches the limit of $F_{\mathrm{av}} \approx$ 0.89 set by the quality of the NV-NV Bell-pair resource state, local two-qubit gate operations, and measurement errors. These results indicate that deterministic two-qubit gate operations across optical quantum networks are within reach.

We next turn to protocols for GHZ state synthesis between four network nodes. We compare three protocols, which are schematically depicted in Fig. \ref{fig:sims}(d). First, we have the `Plain' protocol, the simplest known approach which fuses three distributed Bell pairs into a GHZ state. Second, we have the `Modicum' protocol \cite{de2020protocols,nickerson2014freely}, which extends the Plain protocol by one additional Bell pair creation step between two otherwise idling nodes to realise a single step of entanglement distillation. Finally, we have the `Expedient' protocol introduced by Nickerson \emph{et al.} \cite{nickerson2013topological}, which uses 22 Bell pairs in total to perform additional rounds of distillation, and, in principle, can lead to high-quality GHZ states even in the presence of errors. Note, however, that previous work did not take into account the decoherence of the data qubits during entangling sequences or idling \cite{nickerson2013topological}.

In Fig. \ref{fig:sims}(d), we plot the simulated GHZ state fidelities ($F_\mathrm{state} = \langle \Psi_{\mathrm{GHZ}}|\rho|\Psi_{\mathrm{GHZ}}\rangle$) for each protocol, alongside the average duration to prepare the state. For each scheme, there are two data points, which show the fidelities obtained when considering data qubit lifetimes of $N_{\mathrm{1/e}}$ = 2$\cdot$10$^{3}$ and $N_{\mathrm{1/e}}$ = 2$\cdot$10$^{5}$ respectively. The first observation is that, for all the tested schemes, the improved lifetimes lead to significantly higher fidelities. With the lifetime achieved in this work, the simulated GHZ fidelities for the Plain and Modicum protocols exceed $F=0.5$, witnessing genuine multipartite entanglement of four qubits. 

The differences in performance between the schemes with $N_{\mathrm{1/e}}$ = 2$\cdot$10$^{5}$ give insights into the usefulness of entanglement distillation for the considered parameter regime. First, between the Plain and Modicum protocols, we see that the additional distillation step does improve the fidelity, from F = 0.62 to F = 0.75. However, this fidelity improvement is limited due to the need for an additional SWAP operation (three two-qubit gates). Furthermore, the probabilistic nature of distillation leads to a slight increase in the average duration of the protocol. For the Expedient protocol, we see that it takes significantly longer to successfully generate one GHZ state, and achieves lower fidelities. Additional rounds of entanglement distillation are only beneficial if the fidelity improvements (from distilling imperfect states) exceed the fidelity losses from decoherence of idle qubits. A further increase in $\eta^{*}_{\mathrm{\,link}}$ is required to make use of such protocols.

Together, these results show that the lifetime of the robust quantum-network memory demonstrated in this work is sufficient to demonstrate key distributed quantum computation protocols, such as deterministic two-qubit gates and basic distillation schemes, across optical quantum networks. 

\section{Discussion}

We have demonstrated a robust quantum-network memory based upon a single $^{13}$C spin qubit in isotopically-engineered diamond. Compared with previous work \cite{kalb2018dephasing,pompili2021realization}, the data qubit lifetime during network operation is improved by two orders-of-magnitude. Critically, the data qubit now decoheres more slowly than state-of-the-art entanglement rates between NV centre network nodes \cite{humphreys2018deterministic,pompili2021realization}. Using numerical simulations, we show that the corresponding parameter regime --- with a projected $\eta^{*}_{\mathrm{\,link}}$ $\sim$ 10 --- enables a range of novel protocols for optical quantum networks, including deterministic two-qubit logic and creation of genuinely-entangled four-qubit GHZ states. Additionally, such robust memories would greatly speed up recently demonstrated protocols such as entanglement distillation and entanglement swapping \cite{kalb2017entanglement,pompili2021realization}.

On the path towards reaching the fault-tolerance threshold for large-scale distributed quantum information processing, our simulations show that further improvements in $\eta^{*}_{\mathrm{\,link}}$ are needed. The $^{13}$C decoherence rate $r_{\mathrm{dec}}$ is currently limited by spurious ionisation of the NV centre to NV$^{0}$. Our results show that arbitrary states can be protected while cycling the charge state back and forth with minimal loss of fidelity, suggesting that such events can be mitigated. Further improvements in $r_{\mathrm{dec}}$ are possible by further improving the intrinsic coherence times, and reducing the time needed for the entanglement sequence. Additionally, an improvement of the optical entanglement success probability by a factor $\sim$ 100 is feasible using Fabry-Pérot micro-cavities \cite{riedel2017deterministic,ruf2021resonant}, which would also greatly improve the overall operation speed, as is required for advancing beyond prototype networks and towards technological applications. Together such improvements would yield $\eta^{*}_{\mathrm{\,link}} > 1000$, which is anticipated to be sufficient to realise large-scale error corrected quantum networks \cite{nickerson2013topological}. 

\section{Methods}

\subsection{Sample and hardware setup}

Our experiments are performed on a type-IIa isotopically purified (targeted 0.01\% $^{13}$C) $\langle$100$\rangle$ diamond substrate (Element Six). We address a single NV centre using a cryogenic (4 K) confocal microscope. A solid immersion lens and anti-reflection coating are fabricated to increase optical collection efficiency \cite{hadden2010strongly,robledo2011high,yeung2012anti}. An external magnetic field is applied along the NV symmetry axis using three orthogonal permanent neodymium magnets mounted on linear actuators (Newport UTS100PP). We measure the field vector to be $(\mathrm{B_{x}},\mathrm{B_{y}},\mathrm{B_{z}})$ = $(0.3(1),0.06(8),46.801(1))$ G by spectroscopy of the P1 bath transition frequencies \cite{degen2021entanglement}. 

Resonant optical excitation of NV$^{-}$ at 637 nm (red, Toptica DLPro and New Focus TLB-6704-P) realises high-fidelity spin initialisation (E$_{1,2}$ transitions, := E') and single-shot readout (E$_{\mathrm{y}}$ transition) \cite{robledo2011high}. We measure read-out fidelities of 68.5(2)\% for the bright state ($m_{s} = 0, := \ket{0})$ and 99.3(3)\% for the dark state ($m_{s} = -1, := \ket{1}$), giving $F_{\mathrm{SSRO}}$ = 0.839(3). We employ 515 nm (green, Cobolt MLD) excitation to prepare the NV in NV$^{-}$ and on resonance with the 637 nm lasers \cite{bernien2013heralded}. Finally, for experiments involving the neutral charge state, NV$^{0}$, we use resonant 575 nm (yellow, Toptica DL-SHG Pro) light to realise fast recharging to NV$^{-}$. Through direct current modulation or cascaded acousto-optical modulators, we achieve on/off ratios $>$ 100 dB for all lasers. 

Microwave (MW) driving via a lithographically defined gold stripline enables coherent control between $\ket{0} \leftrightarrow \ket{1}$. Pulses are applied using Hermite envelopes with maximum $\Omega_\mathrm{Rabi}$ $\sim$ 27 MHz, aside from within the entangling primitive, where a multi-tone driving scheme is employed to mitigate heating \cite{supp}. Using a fast microwave switch (TriQuint TGS2355-SM, 40 dB) to suppress electronic noise, we measure long electron spin relaxation ($T_{1} \gg 30$ s), dephasing ($T_{2}^{*}$ = 94(2) $\mu$s) and coherence ($T_{2}$ = 0.992(4) ms) times, the latter of which is limited by the electron spin bath formed by the P1 centres (75 ppb \cite{degen2021entanglement}) which can be reduced in future diamond growth. 

\subsection{Magnetic field stabilisation}

The presence of a number of 6-9 T magnetic field systems in adjacent laboratories leads to slow magnetic field drifts (measured to be $\sim$ 200 mG peak-to-peak across all presented measurements). These drifts are first mitigated using a feedback-loop. We measure the electron spin resonance frequency approximately every 10 minutes, and compensate any deviation to within 3 kHz ($\sim$ 1 mG) of the set-point by moving the magnets. This comes at the cost of slight magnetic field misalignment. If this misalignment becomes too large, or nearby magnetic field systems are being swept too quickly, we observe degradation of the nuclear spin operations. To remove such effects from our results, during $^{13}$C-related data-taking we interleave separate reference measurements of the nuclear spin expectation value after simple state-preparation and measurement. These reference measurements do not contribute to the presented data-sets, but are used to discard measurement runs in the case that the reference measurement falls below 75\% of the calibrated value.

\subsection{Data analysis}

Throughout this text, we do not correct the nuclear spin data for SPAM errors. 

The intrinsic decoherence timescales of the $^{13}$C data qubit, as presented in Fig. 1(d), are fitted as follows:

\renewcommand{\arraystretch}{1.25}
\begin{table}[!h]
    \centering
    \begin{tabular}{|x{2cm}|x{2cm}|x{2cm}|x{2cm}|}
         \hline
         \multicolumn{4}{|c|}{$F(t) = A\,\mathrm{exp}[-(t/T)^{n}]$}\tn \hline
        Metric & $A$ & $T$ (s) & $n$ \tn\hline
        $T_{2,e=\ket{1}}^{*}$ & 0.55(2) & 0.38(1) & 1.6(1) \tn\hline
        $T_{2,e=\ket{0}}^{*}$& 0.46(1) & 0.42(2) & 2.0(2) \tn\hline
        $T_{2,e=\ket{1}}^{N=1}$& 0.51(1) & 1.82(6) & 2.1(2) \tn\hline
        $T_{2,e=\ket{0}}^{N=1}$& 0.52(2) & 1.11(8) & 0.9(1) \tn\hline
        $T_{2,e=\ket{1}}^{N=8}$& 0.58(1) & 2.91(8) & 2.1(2) \tn\hline
        $T_{2,e=\ket{0}}^{N=8}$ & 0.60(2) & 1.62(9) & 1.0(1) \tn\hline
        $T_{1,e=\ket{1}}$ & 0.43(2) & - & - \tn\hline 
        $T_{1,e=\ket{0}}$ & 0.49(2) & 2.8(2) & 0.9(1) \tn\hline
    \end{tabular}
\end{table}

\FloatBarrier

The decoherence timescale for the data qubit during network operation (upper panel of Fig. 2(c)) is fit according to:

\begin{table}[!htbp]
    \centering
    \begin{tabular}{|x{2cm}|x{2cm}|x{2cm}|x{2cm}|}
         \hline
         \multicolumn{4}{|c|}{$F(N) = A\,\mathrm{exp}[-(N/N_{1/e})^{n}]$}\tn \hline
        CR threshold & $A$ & $N_{1/e}$ $\cdot$ 1e5 & $n$ \tn\hline
        0 & 0.49(1) & 1.33(4) & 1.00(5) \tn\hline
        1 & 0.49(1) & 1.80(6) & 1.09(8) \tn\hline
        5 & 0.49(1) & 2.07(8) & 1.09(8) \tn\hline
    \end{tabular}
\end{table} while the fraction of rejected data (lower panel) is fit using:

\begin{table}[!htbp]
    \centering
    \begin{tabular}{|x{2cm}|x{2cm}|x{2cm}|}
         \hline
         \multicolumn{3}{|c|}{$F(N) = 1-A\,\mathrm{exp}[-(N/N_{\mathrm{ion}})]$}\tn \hline
        CR threshold & $A$ & $N_{\mathrm{ion}} \cdot$ 1e5  \tn\hline
        1 & 0.988(4) & 5.03(7) \tn\hline
        5 & 0.91(1) & 3.0(1) \tn\hline
    \end{tabular}
\end{table}

The $^{13}$C nuclear magnetic resonance spectrum for the NV$^{0}$ state (Fig. 3(a)) is described by:
\begin{table}[!htbp]
\centering
\begin{tabular}{|c|c|c|}
\multicolumn{3}{c}{}\\[0em]
\hline
\multicolumn{3}{|c|}{\multirow{2}{*}{\begin{tabular}[c]{@{}c@{}}$F(f) = a - A[\Omega^{2}/2(\Omega^{2} + (f-f_0)^{2})]$ \\ $\cdot[1-\cos{(\sqrt{\Omega^{2}+(f-f_0)^{2}}t)}]$\end{tabular}}} \\
\multicolumn{3}{|c|}{} \\ \hline
$a$ & $A$ & $f_{0}$ (Hz) \\ \hline
0.421(3) & 0.73(2) & 50229.8(1) \\ \hline
\end{tabular}
\end{table}\FloatBarrier where the measured Rabi frequency, $\Omega = 5.4(2)$ Hz, and the chosen pulse duration, $t$ = 92 ms, are fixed parameters.

For the $^{13}$C free induction decay in the NV$^{0}$ state (Fig. 3(b)), we fit:

\begin{table}[!htbp]
    \centering
    \begin{tabular}{|x{2cm}|x{2cm}|x{2cm}|}
        \hline
        \multicolumn{3}{|c|}{\textbf{$F(t) = A\, \mathrm{exp}[-(t/T_{2}^{*})^{n}]$}}\tn \hline
        $A$ & $T_{2}^{*}$ (ms) & $n$ \tn\hline
        0.54(2) & 57(3) & 1.2(1) \tn\hline
    \end{tabular}
\end{table}

\FloatBarrier

Finally, for the free evolution of $\langle X \rangle$ (inset of Fig. 3(b)), we fit:

\begin{table}[!htbp]
    \centering
    \begin{tabular}{|x{2cm}|x{2cm}|x{2cm}|}
        \hline
        \multicolumn{3}{|c|}{$F(t) = A\, \mathrm{exp}[-(t/T_{2}^{*})^{n}]\cdot\cos{(\omega t + \phi)}$}\tn \hline
        $A$ & $\omega$ (Hz) & $\phi$ ($^{\circ}$) \tn\hline
        0.46(1) & 81(1) & -4(3) \tn\hline
    \end{tabular}
\end{table} where $T_{2}^{*}$ = 56 ms and $n$ = 1.2 are fixed to the values found in the previous fit.
\renewcommand{\arraystretch}{1}
\FloatBarrier

\begin{acknowledgments}
We thank M. S. Blok, N. Kalb, M. Pompili, S. L. N. Hermans, H. K. C. Beukers, V. V. Dobrovitski and T. Middelburg for useful discussions. This work was supported by the Netherlands Organisation for Scientific Research (NWO/OCW) through a Vidi grant and as part of the Frontiers of Nanoscience (NanoFront) programme. This project has received funding from the European Research Council (ERC) under the European Union’s Horizon 2020 research and innovation programme (grant agreement No. 852410). We gratefully acknowledge support from the joint research program “Modular quantum computers” by Fujitsu Limited and Delft University of Technology, co-funded by the Netherlands Enterprise Agency under project number PPS2007. This project (QIA) has received funding from the European Union’s Horizon 2020 research and innovation programme under grant agreement No 820445. This work was supported by the Netherlands Organization for Scientific Research (NWO/OCW), as part of the Quantum Software Consortium Program under Project 024.003.037/3368. We gratefully acknowledge SURF (www.surf.nl) for the support in using the supercomputer Cartesius to perform the numerical simulations. S. B. is supported within an Erwin-Schr\"{o}dinger fellowship (QuantNet, No. J 4229-N27) of the Austrian National Science Foundation (FWF).
\end{acknowledgments}

\subsection*{Author contributions}
CEB, SB, RH and THT designed the experiments. CEB performed the experiments. CEB and THT analysed the data. CEB, MJD, SJHL and HPB prepared the experimental setup. SWdB, PFWM and DE performed the network simulations. MM and DJT grew the diamond. CEB and THT wrote the manuscript with input from all authors. THT supervised the project.

\bibliography{Qmem}

\end{document}


	
	\title{Supplementary information for ``Robust quantum network memory \\ based on spin  qubits in isotopically-engineered diamond''}

    \author{C. E. Bradley$^{1,2}$}
	\author{S. W. de Bone$^{1,3}$}
	\author{P. F. W. M\"{o}ller$^{1}$}
	\author{S. Baier$^{1,2}$}
	\author{M. J. Degen$^{1,2}$}
	\author{S. J. H. Loenen$^{1,2}$}
	\author{\\ H. P. Bartling$^{1,2}$}
	\author{M. Markham$^{4}$}
	\author{D. J. Twitchen$^{4}$}
	\author{R. Hanson$^{1,2}$}
	\author{D. Elkouss$^{1}$}
	\author{T. H. Taminiau$^{1,2,\dagger}$}
	
	\affiliation{$^{1}$QuTech, Delft University of Technology, 2628 CJ Delft, The Netherlands}
    \affiliation{$^{2}$Kavli Institute of Nanoscience, Delft University of Technology, 2628 CJ Delft, The Netherlands}
    \affiliation{$^{3}$QuSoft, Centrum Wiskunde \& Informatica, 1098 XG, Amsterdam, The Netherlands}
	\affiliation{$^{4}$Element Six Innovation, Fermi Avenue, Harwell Oxford, Didcot, Oxfordshire, OX110QR, United Kingdom}


	
	\date{\today}
	\maketitle
	
\appendix
\renewcommand\thefigure{\thesection S\arabic{figure}}   
\setcounter{figure}{0} 

\subsection{Nuclear-spin control}

Dynamical decoupling (DD) spectroscopy is used to characterise the nuclear spin environment of the NV centre \cite{taminiau2012detection}. The signal presented in the main text (Fig. 1(c)) is modelled by the interaction of the NV with two individual $^{13}$C spins, taking into account the decoherence of the electron spin during the decoupling sequence \cite{taminiau2012detection}.

The resonance at 44.832 $\mu$s is associated with the single spin characterised as a data qubit in this work, for which we extract hyperfine interaction parameters of $A_{\parallel}$ = $2\pi \cdot$ 80(1) Hz and $A_{\bot}$ = $2\pi \cdot$ 271(4) Hz via Ramsey spectroscopy and DD nutation experiments respectively \cite{boss2016one}. The resonance at 44.794 $\mu$s is, in reality, a sum over a number of spins associated with the $^{13}$C bath. However, the signal is reasonably approximated by a single nuclear spin with $A_{\parallel}$ = $2\pi \cdot$ 0 Hz and $A_{\bot}$ = $2\pi \cdot$ 150 Hz.

We realise universal control over the electron-$^{13}$C system: an electron-nuclear two-qubit gate is implemented following the methods described in Taminiau \emph{et al.} \cite{taminiau2014universal} (CRot duration $\sim$ 25 ms), while selective radio-frequency driving and free precession enable arbitrary $^{13}$C single-qubit gates \cite{bradley2019ten}. Following the methods of Cramer \emph{et al.}, we estimate the two-qubit gate fidelity to be $\sim$ 84\%
\cite{cramer2016repeated}. This is thought to be limited by the electron spin coherence, which is restricted by the P1 spin bath rather than the $^{13}$C bath. Reduction of these impurities through improved diamond growth should enable high fidelity control over a larger number of nuclear spins, as demonstrated in natural abundance $^{13}$C samples \cite{bradley2019ten}.

\begin{figure}[!ht]
    \centering
    \includegraphics[width=1\linewidth]{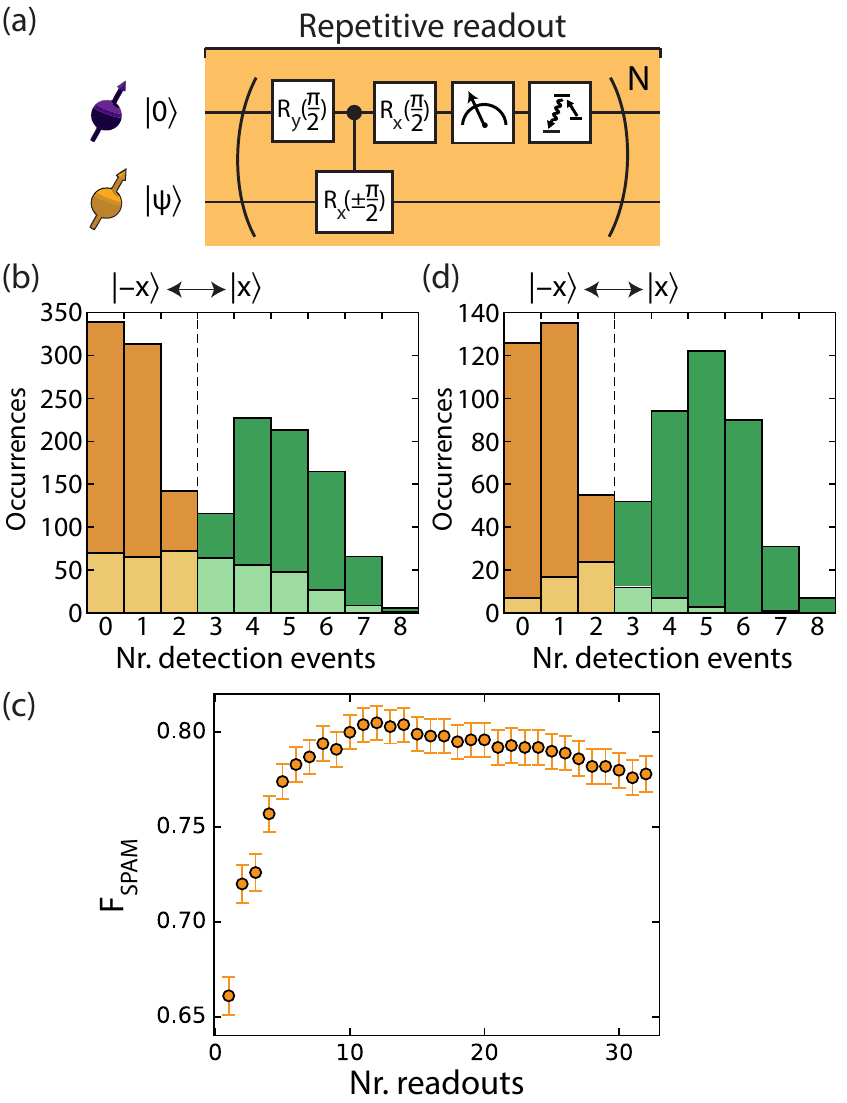}
    \caption{\textbf{Nuclear spin repetitive readout:} (a) Repetitive readout protocol: in each repetition, the nuclear X-basis spin-projection is mapped to the electron spin Z-basis, which is optically read out and then reset. (b) Example histograms obtained when preparing the $\ket{\mathrm{x}}$ and $\ket{\mathrm{-x}}$ states with a single projective measurement, followed by 8 X-basis readouts. Dashed line denotes the optimal state discrimination threshold. Dark (light) colours correspond to populations which are assigned to the targeted (erroneous) state. (c) Optimised state preparation and measurement fidelity as a function of the number of repetitive readouts. (d) Same as (b), but now using the outcomes of 8 prior readouts ($\geq$5 read-outs with $\geq$1 photon for $\ket{\mathrm{x}}$ initialisation, no photons in any of the 8 readouts for $\ket{\mathrm{-x}}$ initialisation) as an additional initialisation step.} 
    \label{fig:repRO}
\end{figure}

As the fidelity of the nuclear spin readout is limited by the electron spin coherence, it is possible to improve this readout fidelity by repeated measurements, as has previously been achieved for single $^{13}$C nuclear spins under ambient conditions \cite{maurer2012room,dreau2013single,liu2017single}. We implement the repetitive readout protocol shown in Fig. \ref{fig:repRO}(a). As the electron-nuclear coupling is weak --- $A_{\parallel} \ll 1/\tau_{\mathrm{RO}}$, for $\tau_{\mathrm{RO}}$ the combined readout and repumping time --- electron spin flips during the optical readout induce minimal $^{13}$C dephasing. We can thus repeat the mapping and readout procedure a number of times, from which we acquire histograms as exemplified in Fig. \ref{fig:repRO}(b). 

The histograms after preparation of $\ket{\mathrm{x}}$ and $\ket{\mathrm{-x}}$ respectively show two distinct populations. The state preparation and measurement process is quantified by $F_{\mathrm{SPAM}} = \frac{1}{2}(F_{\mathrm{x|x}} + F_{\mathrm{-x|-x}})$, where $F_{i|j}$ is the probability to assign the spin state $\ket{i}$ after attempting initialisation in $\ket{j}$. In Fig. \ref{fig:repRO}(c), we plot $F_{\mathrm{SPAM}}$ as a function of the number of repeated readouts, in each case using the optimal threshold to discriminate the two states. At an optimum number of 12 readouts, $F_{\mathrm{SPAM}}$ reaches 80.4(9)\%. In all nuclear spin data shown in the main text and in the remainder of this supplement, we utilise N = 8 readout repetitions ($F_{\mathrm{SPAM}}$ = 79.4(9)\%) as a trade-off between optimal fidelity and measurement run-time. The associated histogram is shown in Fig. \ref{fig:repRO}(b). 

$F_{\mathrm{SPAM}}$ is predominantly limited by two mechanisms. First, the measurement sequence still induces some decoherence of the nuclear spin --- as can be seen by the decrease in fidelity for $>$12 read-outs --- reducing the extractable information. This effect would likely be mitigated by improvement of the two-qubit gate. Second, we prepare the $\ket{\mathrm{x}}$ and $\ket{\mathrm{-x}}$ states using a single projective measurement \cite{cramer2016repeated}. Repetitive measurements can also be used to initialise the spin: in Fig. \ref{fig:repRO}(d) we additionally show the outcome of 8 repetitive readouts, conditioned on the outcome of 8 prior readouts. We find an achievable $F_{\mathrm{SPAM}}$ = 91(1)\%, but do not employ it in this work due to the additional time overhead of $>$0.5 s per experimental shot.

\subsection{$^{13}$C spin-bath echo}

In the main text, we report the nuclear spin dephasing timescale associated with nuclear spin bath dynamics to be $T_{2,^{13}\mathrm{C-bath}}^{*}$ = 0.66(3) s. To identify this timescale, we prepare the electron in the $\ket{0}$ state and perform a spin-echo at the $^{13}$C Larmor frequency ($\Omega_\mathrm{Rabi}$ = 120 Hz). Such a pulse inverts the local nuclear spin environment, decoupling it from other quasi-static noise sources but leaving the $^{13}$C-$^{13}$C interactions unperturbed.

\begin{figure}[!htb]
    \centering
    \includegraphics[width=0.9\linewidth]{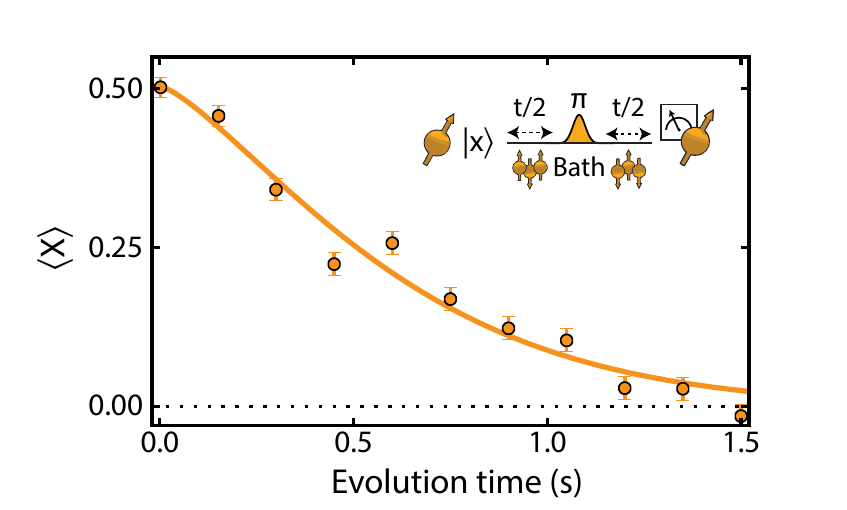}
    \caption{\textbf{$^{13}$C spin-bath echo:} Measured nuclear spin dephasing when applying a spin-echo at the $^{13}C$ Larmor frequency. The electron is initialised in the $\ket{0}$ state, such that the echo pulse simultaneously inverts the addressed $^{13}$C along with the surrounding nuclear spin environment. The data are fit with an exponential decay, $f(t) = A\, \mathrm{exp}[-(t/T_{2})^{n}]$, for which we find $T_{2,^{13}\mathrm{C-bath}}^{*}$ = 0.66(3) s and $n$ = 1.3(1).} 
    \label{fig:bathecho}
\end{figure}

In Fig. \ref{fig:bathecho} we present the measured data and experimental sequence.

\begin{figure}[!ht]
    \centering
    \includegraphics[width=\linewidth]{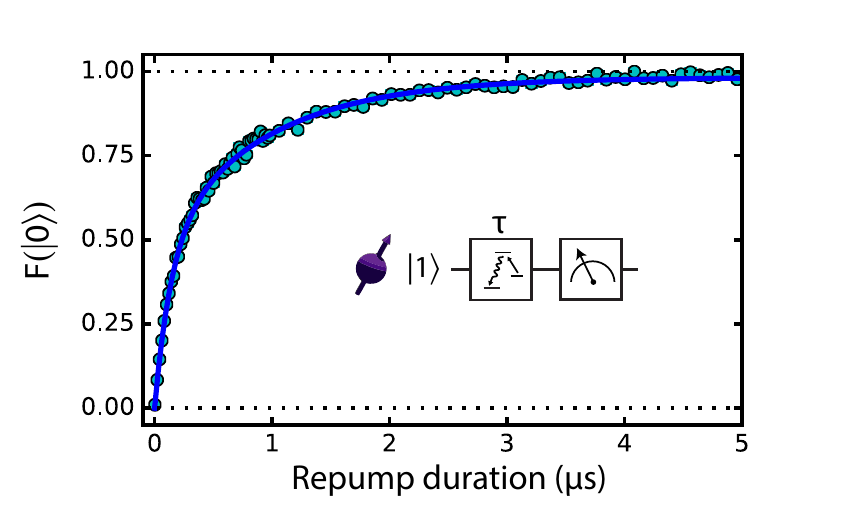}
    \caption{\textbf{Electron spin reset:} After initialisation in $\ket{1}$, the population in $\ket{0}$ is plotted as a function of the repumping duration (30 nW, E'). A delay of 10 $\mu$s between turning off the spin-pumping laser and turning on the readout laser ensures that minimal population resides in the metastable singlet states \cite{kalb2018dephasing}. The data are well described by a two-timescale phenomenological model: $f(t) = A (1-\mathrm{exp}[-t/\tau_{1}]) + B(1-\mathrm{exp}[-t/\tau_{2}])$, for which we find $A$=0.480(15), $B$=0.503(14), $\tau_{1}$=142(5) ns, $\tau_{2}$=905(31) ns.} 
    \label{fig:repump}
\end{figure}

\subsection{Electron spin initialisation} 

Electron spin initialisation constitutes a large fraction of the remote entanglement primitive duration \cite{pompili2021realization}. Therefore, faster initialisation rates are desirable. This can be achieved by using higher laser powers, such that the NV centre is more frequently excited. However, higher laser powers also increase the ionisation rate \cite{robledo2010control,siyushev2013optically}. Furthermore, as initialisation occurs primarily via the NV$^{-}$ intersystem crossing \cite{kalb2018dephasing}, the $\sim$ 370 ns lifetime of the singlet state, alongside the finite branching ratio from this state to the targeted $m_{s}=0$ state, limit the extent to which higher laser powers increase the reset speed.

As a trade-off between these factors, we use 30 nW of E' excitation for the repumping process. This leads to initialisation fidelities $\geq$98\% in a duration of 5 $\mu$s. In Fig. \ref{fig:repump} we plot the measured reset dynamics for this power. 

\subsection{Weak electron-spin pulses}

When implementing large numbers of repetitions of the remote entanglement protocol using strong microwave pulses, the mean number of detected photons during single-shot read-out was observed to decrease. This is attributed to heating of the sample. Note that heating is problematic in the context of optical entanglement generation, because a reduction of the photon emission decreases the success probability, and the photon indistinguishability also decreases with temperature \cite{bernien2013heralded}.

The applied heat-load can be reduced by decreasing the MW Rabi frequency, but this is complicated by the presence of the nitrogen nuclear-spin. The electron-nitrogen hyperfine coupling, $A_{\parallel}{S_{z}}{I_{z}}$ ($A_{\parallel} = 2\pi \cdot 2.18$ MHz) causes a three-fold splitting of the transition frequency. For this reason, we usually employ strong, spectrally-broad Hermite pulses (Rabi frequency $\sim$ 27 MHz) for high-fidelity ($\gg 99\%)$ manipulation of the electron spin, independent of the nitrogen spin state. 

While prior initialisation of the nitrogen spin in a chosen state can relax this requirement \cite{kalb2016experimental}, the large excited-state hyperfine coupling causes rapid depolarisation of the nuclear spin under repeated optical reset of the electron-spin, ruling out this method in the presented experiments \cite{pfaff2014unconditional}.

An alternative approach is to apply three separate drive tones, each addressing one of the nitrogen hyperfine levels. High-fidelity inversion is realised if the Rabi frequency of each tone, $\Omega$, is large relative to the transition linewidth, which is described by a Gaussian distribution with $\sigma_{f} = 1/(\sqrt{2}\pi T_{2}^{*}) = 4.5$ kHz. Conversely, $\Omega$ should be small relative to the detuning between the transitions $A_{\parallel}$, such that AC-Stark frequency shifts remain small. We apply square pulse envelopes with $\Omega$ = 92 kHz for each tone. The integrated power over the multi-tone pulse is $\sim 80$ ($\sim 40$) times reduced for the $\pi$ ($\pi/2$) pulse when compared to the strong Hermite pulses (Rabi frequency: $\pi$-pulse $\sim$ 27 MHz, $\pi/2$-pulse $\sim$ 16 MHz), enabling the repeated application of hundreds of thousands of MW pulses without observation of heating effects.

We numerically evaluate the electron spin dynamics using the QuTip python package \cite{johansson2012qutip}. Making the secular and rotating wave approximations, and transforming to the interaction picture $\omega_{1} \hat{S}_{z}$, with $\omega_{1} = D + \gamma_{e} B_{z}$, the Hamiltonian describing the evolution of the driven electron-nitrogen system is:

\begin{equation}\begin{split}
    H' = &\ 2D \ket{2}\bra{2} + Q \hat{I}^{2}_{z} + \gamma_{N} B_{z} \hat{I}_{z} + A_{\parallel} \hat{S}_{z}\hat{I}_{z} \\
    & + \frac{1}{\sqrt{2}}\Omega (\cos(\phi_1)\hat{S}_{x} - \sin(\phi_1)\hat{S}_{y}) \\
    & + \frac{1}{\sqrt{2}}\Omega (\cos(A_{\parallel} t + \phi_2)\hat{S}_{x} - \sin(A_{\parallel} t + \phi_2)\hat{S}_{y}) \\
    & + \frac{1}{\sqrt{2}}\Omega (\cos(-A_{\parallel} t + \phi_3)\hat{S}_{x} - \sin(-A_{\parallel} t + \phi_3)\hat{S}_{y}).
\end{split}\end{equation}
Here, $\hat{S}_{i}$ and $\hat{I}_{i}$ are the electron and nuclear spin-1 operators. D is the zero-field splitting, Q is the quadrupole splitting, $\gamma_{e}$ and $\gamma_{N}$ are the electron and nitrogen gyromagnetic ratios, and $B_{z}$ is the external magnetic field component parallel to the NV axis. $\phi_{i}$ are the phases of the individual drive fields, while $\ket{2}\bra{2}$ is the projector on the state $\ket{2}$ of the electron spin: this term is a detuning of the non-driven $m_{s} = +1$ electron spin transition. We have additionally assumed that the three drive tones are x-polarised.

\begin{figure}[!t]
    \centering
    \includegraphics[width=\linewidth]{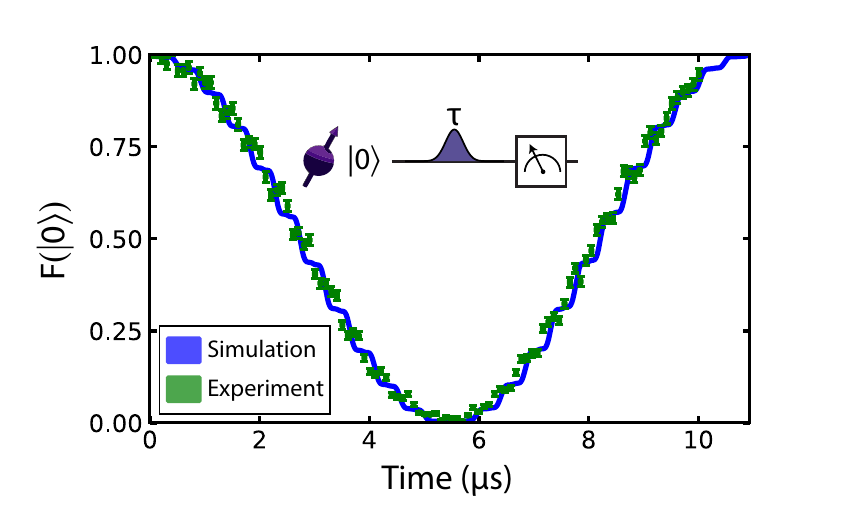}
    \caption{\textbf{Weak microwave pulses:} Electron spin dynamics under application of the three-tone microwave pulse.  The solid line (circles) show the simulated (measured) dynamics after initialisation of the electron spin in $\ket{0}$. A $\pi$-pulse is realised after 5.5 $\mu$s.}
    \label{fig:weakpulse}
\end{figure}

In Fig. \ref{fig:weakpulse} we show the simulated and measured electron spin dynamics. A high-fidelity $\pi$-pulse is realised after $\sim$ 5.5 $\mu$s of driving. The observed modulations arise due to slight relative AC-Stark shifts. Each drive tone induces a frequency shift approximately given by $\delta_{\mathrm{AC-Stark}}\,\sim\,\Omega^{2}/2\Delta$, where $\Delta$ is the difference between the drive frequency and each detuned transition frequency. For the $m_{I} = 0$ transition, the shifts from the off-resonant tones are cancelled to first order. However, for the $m_{I} = \pm\,1$ transitions, there is a frequency shift of $\sim \pm 2$ kHz. 

We note that the parameters used here are not suitable for natural abundance samples, ($T_{2}^{*} \sim 5 \, \mu$s, $1/(\sqrt{2}\pi T_{2}^{*}) \sim 45$ kHz), where the broader intrinsic linewidth leads to significant infidelity (simulated to be $\sim 33\%$ by Monte Carlo sampling, compared with $\sim 0.6\%$ for the present scenario). However, heating can alternatively be addressed by improved device engineering in future work, for example by fabricating the stripline using superconducting materials. This would enable the use of strong microwave pulses throughout the experiments and remove the time overhead required for weak pulses.

\subsection{Quantum network memory performance using strong pulses}

As discussed in the above section, the use of strong microwave pulses during the remote entanglement primitive led to the observation of heating effects. For this reason, we implemented the weak pulse scheme to ensure that optical processes were unaffected and thus would be compatible with remote entanglement generation. Nevertheless, we here study the data qubit decoherence when using strong pulses, which enables a shorter duration per entangling attempt. 

\begin{figure}[!t]
    \centering
    \includegraphics[width=0.85\linewidth]{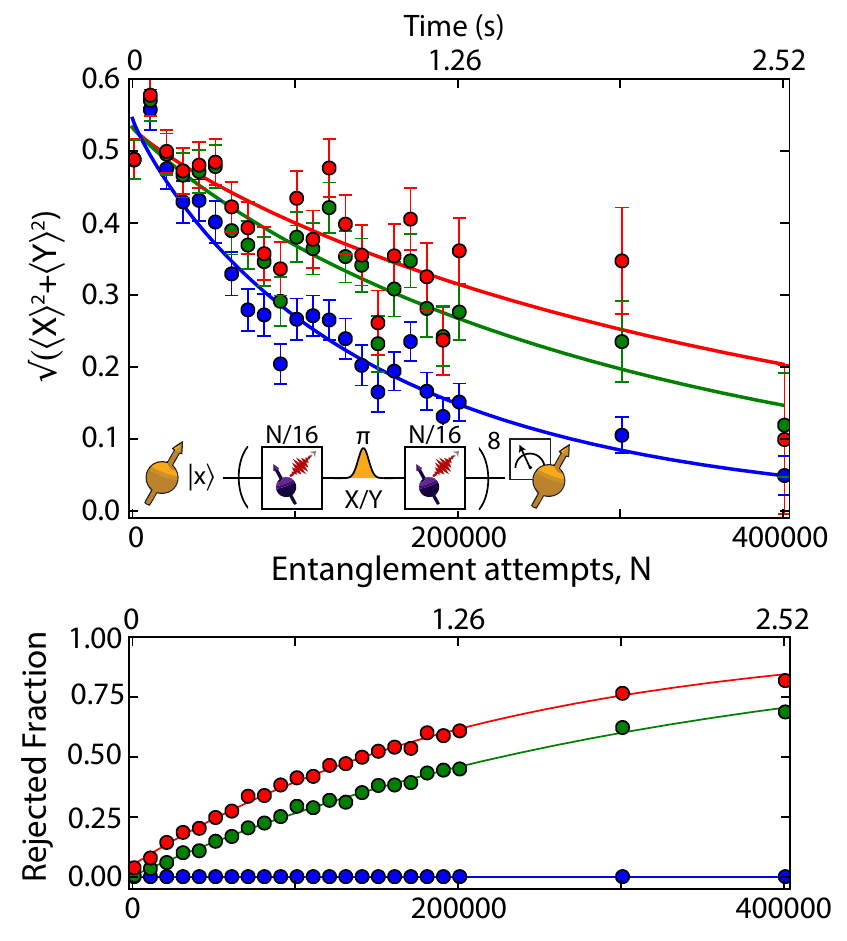}
    \caption{\textbf{Data qubit robustness using strong electron pulses:} Nuclear spin coherence as a function of the number of applied remote entangling primitives. Blue, green, red lines correspond to post-selection using a CR-check performed after the experiment, with thresholds of 0, 1, 5 respectively. The lower panel shows the corresponding fraction of rejected data.} 
    \label{fig:shortLDE}
\end{figure}

The entangling primitive used here is identical to that used in the main text ($t_{\mathrm{r}}$ = 5 $\mu$s, $t'$ = 200 ns, $t_{\mathrm{l}}$ = 1 $\mu$s), but with the weak MW $\pi$/2 pulse ($t_{\mathrm{MW}}$ = 2.8 $\mu$s) now replaced by a Hermite $\pi$/2 pulse ($t_{\mathrm{MW}}$ = 100 ns). The total duration is thus reduced from 9 $\mu$s to 6.3 $\mu$s. We measure the decay of the nuclear coherence, $(\langle X \rangle ^{2} + \langle Y \rangle ^{2})^{1/2}$, as a function of the number of entangling primitives after initialisation in $\ket{\mathrm{x}}$. We again interleave one round of nuclear spin XY8 decoupling pulses. In Fig. \ref{fig:shortLDE} we plot the results, including different post-selection thresholds for the post-measurement CR checks as in the main text. Importantly, the rejected fraction of events is similar to that observed when using the weak pulse scheme, suggesting that optical dynamics are relatively unchanged.

When compared with the data presented in the main text, we observe an increase in data qubit robustness. With the most stringent filtering, we find a decay time of 419(96)$\cdot$10$^{3}$ attempts (2.6(6) s). This compares favourably with the weak pulse case, where the equatorial states \{$\ket{\mathrm{x}}$,$\ket{\mathrm{y}}$,$\ket{\mathrm{-x}}$,$\ket{\mathrm{-y}}$\} have a mean decay time of 190(8)$\cdot$10$^{3}$ attempts (1.8(1) s). 

\subsection{Simulated nuclear spin dephasing due to the remote-entanglement protocol}

\begin{figure}[!ht]
    \centering
    \includegraphics[width=\linewidth]{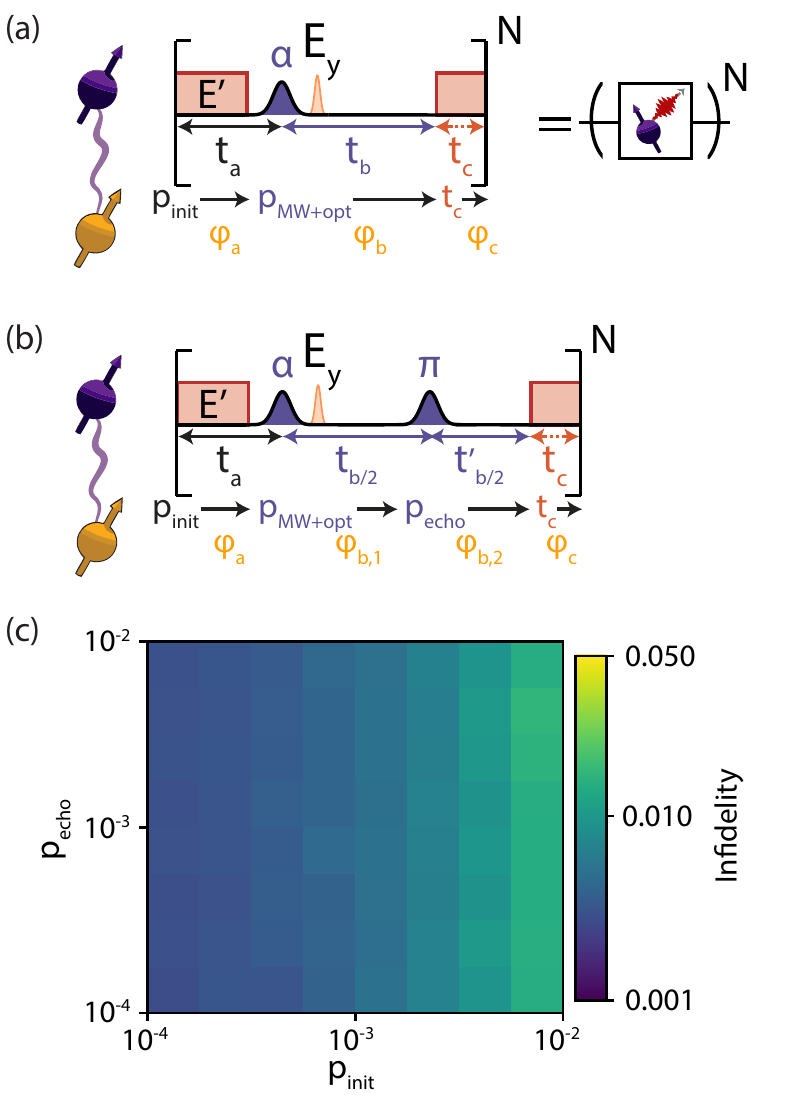}
    \caption{\textbf{Numerical analysis of nuclear spin dephasing:} (a) Sketch of the primitive remote entanglement unit used in experiment as formulated for Monte Carlo simulation. The nuclear spin (yellow) acquires spurious phases according to the dynamics of the electron spin (purple) and the $\hat{S}_{z}\hat{I}_{z}$ coupling $A_{\parallel}$. This phase is broken down into contributions from three imperfect operations, determined by initialisation infidelity ($\phi_{a}$, parameterised by $p_{\mathrm{init}}$), microwave inversion ($\phi_{b}$, parameterised by $p_{\mathrm{MW}}$ and $p_{\mathrm{opt}}$) and stochastic reset time ($\phi_{c}$, parameterised by $t_{c}$). (b) Sketch of the primitive remote entanglement unit when incorporating an additional electron spin-echo pulse (with failure possibility parameterised by $p_{\mathrm{echo}}$). (c) Results of Monte Carlo simulations for the scenario given in (b), after application of $10^{6}$ entangling primitives. $\alpha$ = $\pi$/2, $p_{\mathrm{MW}}$ = 0.5, $p_{\mathrm{opt}}$ = 0.01, $t_{a}$ = 5.1 $\mu$s, $t_{b}$ = 1.3 $\mu$s and $\langle t_{c} \rangle$ = 200 ns. Results are averaged over 1000 samples.} 
    \label{fig:MCsim}
\end{figure}

In this section we numerically analyse the nuclear-spin dephasing induced by repeated remote entanglement attempts. Our analysis is based upon Monte Carlo simulations, adapted from the work of Kalb \emph{et al.} \cite{kalb2018dephasing}. Here, we will neglect the intrinsic dephasing timescales ($T_{2}^{*}$ and $T_{2}$) of the nuclear spin, rather focusing on the limits imposed by the remote entanglement sequence alone. Further, we neglect the role played by ionisation. We will consider two sequences, the first following that implemented experimentally, and the second in which an additional electron spin-echo is incorporated within each attempt, see Figs. \ref{fig:MCsim}(a,b). 

To build up a statistical distribution for the phase acquired by the nuclear spin after repeated application of the sequence, we generate a number of samples, $K$, each of $N$ trials. 

In each trial, a random number $0 \leq \mathrm{R}_{\mathrm{init}} \leq 1$ is used to instantiate the NV electron in a given spin projection, $\ket{s}\in\,\{-1,0,+1\}$, each with probability $p_{\ket{s}}$. The $p_{\ket{s}}$ are parameterised by a chosen initialisation error, $p_{\mathrm{init}}$, such that $p_{\ket{0}} = (1-p_{\mathrm{init}})$ and $p_{\ket{-1}} = p_{\ket{+1}} = p_{\mathrm{init}}/2$. 

The nuclear spin is now evolved for a time, $t_{\mathrm{a}} = (t_{\mathrm{p}} - \langle t_{c} \rangle + t_{\mathrm{s}})$, where $t_{\mathrm{p}}$ is the total optical pumping duration, $\langle t_{c} \rangle$ is the mean time to initialise (`reset') the spin, and $t_{\mathrm{s}}$ is the spacing between the end of the repump laser pulse and the centre of the `$\alpha$' microwave pulse (used to prepare the electron superposition for spin-photon entanglement). The stochastic nature of $t_{c}$ is accounted for below. A phase, $\phi_{a}$ is acquired according to the initial state, at the difference frequency $\delta\omega = A_{\parallel}s$.

Next, the `$\alpha$' microwave pulse is applied, the effect of which is treated as instantaneous. While throughout this work we set $\alpha$ = $\pi/2$, for which the sequence induces maximal dephasing noise on the data qubit, $\alpha$ is generally a free parameter which determines a rate-fidelity trade-off for remote spin-spin entanglement in the single-click entangling protocol \cite{humphreys2018deterministic,pompili2021realization}. Another random number, $0 \leq \mathrm{R}_{\mathrm{MW}} \leq 1$, is used to determine whether the electron spin is flipped between $\ket{0} \leftrightarrow \ket{-1}$ ($\ket{+1}$ is assumed to be unperturbed). The inversion is applied if $\mathrm{R}_{\mathrm{MW}} < p_{\mathrm{MW}}$, the inversion probability of the chosen microwave pulse.

Immediately after the microwave pulse, we apply the effect of the optical $\pi$-pulse used to create spin-photon entanglement. A third random number, $0 \leq \mathrm{R}_{\mathrm{opt}} \leq 1$, is used to determine whether the electron spin is flipped from $\ket{0} \rightarrow \{\ket{-1},\ket{+1}\}$, each with equal probability. Again, the inversion is applied if $\mathrm{R}_{\mathrm{opt}} < p_{\mathrm{opt}}$, which we set to be 0.01, according to the estimated spin-flip probability of the E$_{\mathrm{x,y}}$ transitions \cite{kalb2018dephasing}. Note that the probability of this spin-flip is much smaller than that induced by the microwave pulse, justifying the omission of the optical pulses in the presented experiments.

In the case that we do not incorporate the electron spin-echo (Fig. \ref{fig:MCsim}(a)), the nuclear spin now simply acquires phase for a time $t_{b}$ ($\phi_{b}$), defined as the time between the centre of the microwave pulse and the onset of optical pumping. This time period allows for the decision logic for success or failure of the remote entanglement protocol (detection or loss of the optical photon). For short range (metre-scale) experiments, this timescale is typically $\sim 1\, \mu s$, limited by the current electronics used. The dephasing associated with $t_{b}$ is dominant in this sequence: for $\alpha$ = $\pi/2$, there is an extended period of time during which the electron has an approximately 50\% chance to be in either the $\ket{0}$ or $\ket{-1}$ state. 

Finally, we account for the stochastic optical reset. In the case that the electron spin is in $\ket{-1}$ or $\ket{+1}$ at the onset of spin-pumping, a finite duration is required to prepare $\ket{0}$ for the start of the next repetition. A reset time, $t_{c}$, is sampled from the phenomenological two-timescale function fitted in Fig. \ref{fig:repump}, which determines the final evolution time before the end of the trial, during which time a phase $\phi_{c}$ is acquired.

Summing these phases, we find the total phase acquired by the nuclear spin for the $i$-th trial of the $k$-th sample: $\phi_{i,k}$. Repeating this process for $N$ trials, we calculate the cumulative phases $\Phi_{n,k} = \sum_{i=1}^{n} \phi_{i,k}$ for $n = \{1,...,N\}$. 

For each sample, we can then calculate the state fidelity after each $n$ trials:
\begin{equation}
    F_{n,k} = \frac{1}{2} + \frac{1}{2} \cos(\Phi_{n,k}-\overline{\Phi_{n}})
\end{equation}

where $\overline{\Phi_{n}}$ is the mean phase after $n$ trials, averaged over all samples. Finally, we find the characteristic $n$ for which $\cos(\Phi_{n,k}-\overline{\Phi_{n}})$ falls below $1/e$.

As the dephasing associated with $t_{b}$ is dominant, 
we also consider the case in which it is possible to incorporate an additional spin-echo into the sequence (for example, because heating is mitigated by device engineering), see Fig. \ref{fig:MCsim}(b). In this case, the nuclear spin acquires phase for a time $t_{b}$/2 ($\phi_{b,1}$), before the echo is applied, and then for a further time $(t'_{b}/2 + \langle t_{c} \rangle) = t_{b}$/2 ($\phi_{b,2}$). In case of successful application of the echo, $\phi_{b,1} + \phi_{b,2}$ = 0. We account for a finite failure probability of the echo pulse using another random number, $0 \leq \mathrm{R}_{\mathrm{echo}} \leq 1$. If $\mathrm{R}_{\mathrm{echo}} > p_{\mathrm{echo}}$, the electron spin is flipped between $\ket{0} \leftrightarrow \ket{-1}$.

We consider two regimes. First, for the sequence excluding the spin-echo, we choose parameters that are as least as pessimistic as used in the experiments in the main text. We have $t_{a}$ = 6.1 $\mu$s, $t_{b}$ = 2.4 $\mu$s, $\langle t_{c} \rangle$ = 535 ns (following the two-timescale function fitted in Fig. \ref{fig:repump}), $p_{\mathrm{init}}$ = 0.03, $p_{\mathrm{MW}}$ = 0.5, $p_{\mathrm{opt}}$ = 0.01. We find that after application of $10^{6}$ entangling primitives, the expected fidelity with an initial superposition state is 0.776(4), averaged over $K$ = 4000 samples. These calculations indicate that dephasing due to the entangling sequence is not limiting in the experiments.

We now turn to the case in which the echo is incorporated. We assume that strong MW pulses can be employed, enabling $t_{a}$ = 5.1 $\mu$s, $t_{b}$ = 1.3 $\mu$s. Further, we assume that faster spin reset can be achieved, as has been previously demonstrated with higher powers \cite{reiserer2016robust,kalb2018dephasing}: setting $\langle t_{c} \rangle$ = 200 ns (following a single exponential timescale). We set $p_{\mathrm{MW}}$ = 0.5, $p_{\mathrm{opt}}$ = 0.01. We now perform a 2D sweep of the fidelity after $10^{6}$ trials as a function of $p_{\mathrm{init}}$ and $p_{\mathrm{echo}}$. The results are plotted in Fig. \ref{fig:MCsim}(c). Even with 1\% errors in the initialisation and echo operations, a fidelity of 0.975(1) is achieved. Note that the errors associated with initialisation are dominant over those associated with echo failure for these timings. This is because, first, $t_{a} > t_{b}$, and, second, this process can erroneously prepare the electron $\ket{s} = +1$ state. In particular, during the time $t_{b}$, the memory ideally evolves at the average frequency associated with the electron $\ket{s} = \{-1,0\}$ states. Instances in which the $\ket{s} = +1$ state occurs lead to $t_{b}$ evolution at a frequency difference 3 times larger than cases in which the echo fails (which result in $t_{b}$ evolution with either $\ket{s} = 0$ or $-1$).

\subsection{Controlled charge-state switching of the NV centre}

Throughout the experiments shown in this work, we make use of two-laser probe measurements (`CR checks') to detect the NV$^{-}$ charge state \cite{bernien2013heralded}. During these checks, weak resonant light is applied on the NV$^{-}$ E' (8 nW) and E$_{\mathrm{y}}$ (1.5 nW) transitions for 150 $\mu$s, during which we record all photon counts. By setting a threshold on the number of detected photons, it is possible to herald that the NV is in the negative charge-state, and that the spin-pumping and readout lasers are resonant. In the case of low counts a strong green pulse (515 nm, 30 $\mu$W, 50 $\mu$s) is used to scramble the local charge environment, before the check is repeated.

To investigate the effect of NV$^{-}$ ionisation on the nuclear spin data qubit, we require fast switching of the NV charge state along with efficient detection. In this section we design and implement a protocol which enables such studies.

\begin{figure*}
		\includegraphics[width=1.\linewidth]{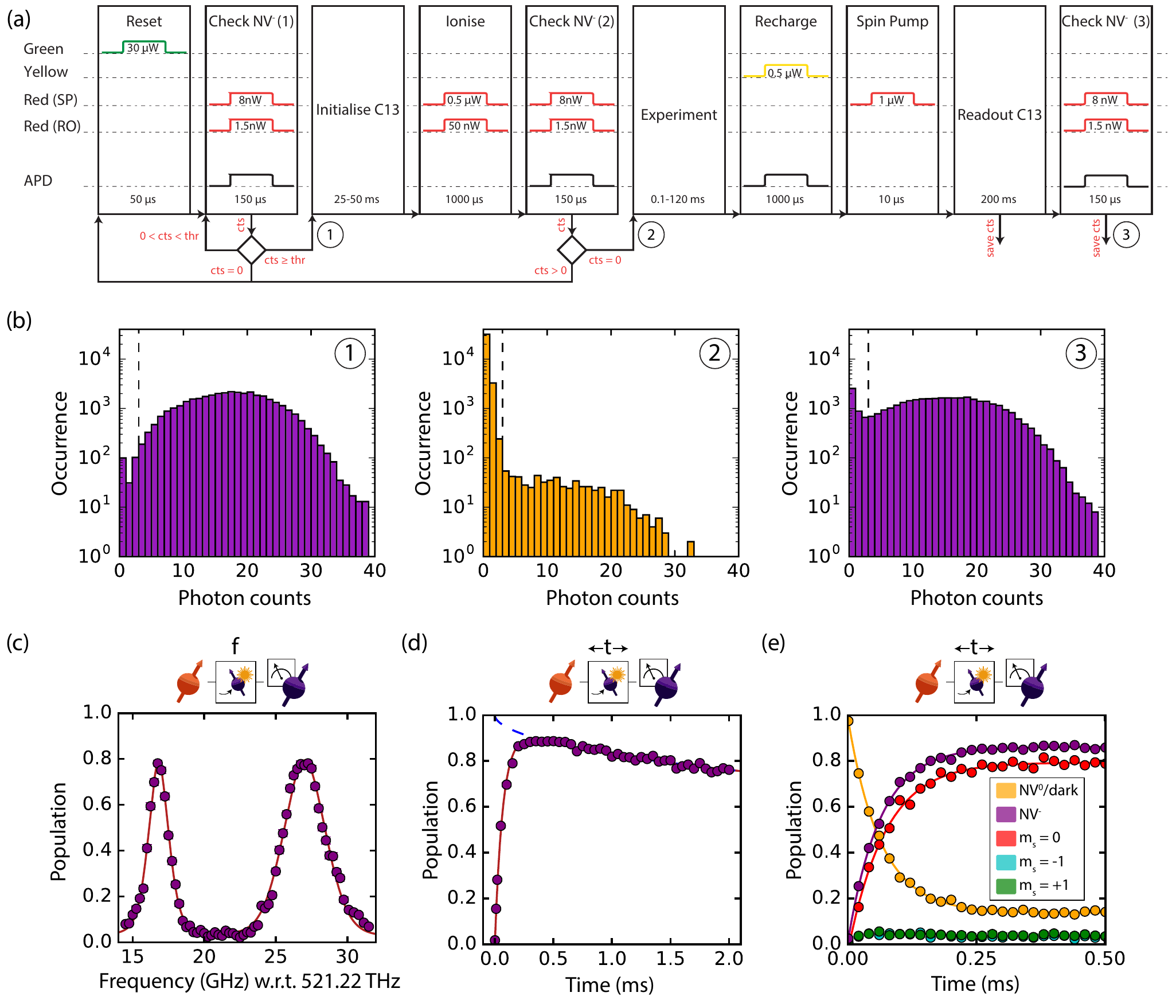}%
		\caption{\label{fig:chargeswitch} \textbf{Charge-state switching:} (a) Flow diagram for the charge-state switching protocol, showing the decision logic used for bright- and dark-state heralding and optional post-selection. (b) Distribution of photon-counts from CR checks performed at the points (1), (2), (3) indicated in panel (a). (c) NV$^{-}$ population as a function of the recharging pulse frequency, using a 1 ms pulse of 500 nW. Solid line is a fit to the sum of two Voigt profiles. (d) NV$^{-}$ state population as a function of the recharging pulse duration, at 521.247 THz and 1.1 $\mu$W. Solid line is a fit to $F(t) = (1-\mathrm{exp}[-t/\tau_{1}])\cdot(\mathrm{exp}[-(t/\tau_{2})^{n}]$, from which we find $\tau_{1}$ = 71(2) $\mu$s, $\tau_{2}$ = 22(4) ms and $n$ = 0.54(3).
		Dashed line shows the decay component, characterised by $\tau_{2}$. (e) Populations in the NV$^{0}$/dark state, and in the differing spin-projections of the NV$^{-}$ state as a function of the recharging pulse duration, again at 521.247 THz and 1.1 $\mu$W. Solid lines are fits to $F(t) = a + A\,\mathrm{exp}[-(t/\tau)^{n}]$. We find $\tau$ = \{73(7),63(1),63(1)\} $\mu$s, $n$ = \{1.0(1),1.00(1),1.00(1)\}, for the $m_{s}=0$, NV$^{0}$ and NV$^{-}$ populations respectively.}
\end{figure*}

The protocol is shown in Fig. \ref{fig:chargeswitch}(a). We begin with a CR check, which we use to prepare the system in NV$^{-}$ (Check NV$^{-}$ (1)). At this point, we can utilise the previously described universal control to prepare the nuclear spin in a chosen state. Next, we apply a strong two-laser ionisation pulse. 50 nW of E$_{\mathrm{y}}$ and 500 nW of E' excitation are used, which are chosen to maintain predominantly NV$^{-}$ $m_{s} = 0$ projection while attempting ionisation. After this pulse, a second CR check is performed (Check NV$^{-}$ (2)). We proceed with the measurement only if this check returns 0 photon counts, which occurs with $\sim$ 10\% (40\%) probability after a 300 $\mu$s (1 ms) ionisation pulse. This outcome heralds the system in a dark state of the NV, which can be associated with either NV$^{0}$ or a significant detuning of the NV$^{-}$ transitions due to spectral diffusion. 

We now perform the desired experiment on the nuclear spin, either allowing free evolution or applying RF driving. Next, we apply a strong yellow (575 nm) recharging pulse, using 500 nW for main text Figs. 3(a,b), and 1100 nW (the maximum attainable) for Fig. 3(c) to increase the recharging probability. After the recharging pulse, we apply spin-pump light to prepare the NV$^{-}$ $m_{s} = 0$ state and subsequently read-out the nuclear spin. Finally, we perform one more CR check (Check NV$^{-}$ (3)). The outcome of this check allows for post-selection of the nuclear spin read-out data upon finding the NV in the negative-charge state (retaining 82\% of cases with the 575 nm pulse parameters used for Figs. 3(a,b): 1 ms, 500 nW, 85\% for Fig. 3(c): 0.5 ms, 1100 nW). For post-selection, we discard data which lies close to the bright/dark discrimination threshold (see below). We only consider data for which the NV began in the negative charge-state ($>$ 13 counts in (1)), transferred to a dark state (0 counts in (2)), and then returned to the negative charge-state via the recharging pulse ($\geq$ 5 counts in (3)). 

In Fig. \ref{fig:chargeswitch}(b) we show example photon-count distributions from CR checks performed after preparing NV$^{-}$ (1), after heralding the dark state (2), and after applying a 1.1 $\mu$W, 0.5 ms recharging pulse (3). The mean number of photons detected in each case is 16.9, 0.3, and 13.7 respectively, clearly demonstrating the ability to switch between the bright and dark states. However, the slight reduction of mean counts for case (3) compared to case (1) indicates that the recharging process does not achieve unity success. We now study this process more carefully.

We first quantify the charge-state read-out fidelity. Considering the data shown in panels (1) and (2) of Fig. \ref{fig:chargeswitch}(b), we maximise the quantity $F_{\mathrm{charge}} = \frac{1}{2}(F_{\mathrm{b|b}} + F_{\mathrm{d|d}})$, where $F_{i|j}$ is the probability to assign the `bright' (b, NV$^{-}$) or `dark' (d, NV$^{0}$ or off-resonant) state after attempting initialisation in that state. With a threshold of 3 counts (assign bright if $\geq$ 3 photons detected), we find $F_{\mathrm{charge}}$ = 98.81(4)\%. 

In Fig. \ref{fig:chargeswitch}(c) we plot the NV$^{-}$ population after first preparing the dark state and then applying a 1 ms, 500 nW recharging pulse, for which we sweep the pulse frequency. The observation of two transitions corresponding to the spin-orbit branches of NV$^{0}$ confirms that this is the resonant recharging process \cite{barson2019fine,baier2020orbital}. Here, the broader line observed for the higher-frequency resonance (lower spin-orbit branch of NV$^{0}$) is likely due to the polarisation of the yellow laser --- which is fixed at an arbitrary value in these experiments --- being biased towards linear horizontal \cite{baier2020orbital}. Optimal recharging is achieved when working at the central frequency for this branch, here 521.247 THz. Note that the lifetime of the NV$^{0}$ orbital-state is short even at 4 K (sub-$\mu$s), and so it is sufficient to only address a single spin-orbit branch.

We now turn to the temporal dependence. In Fig. \ref{fig:chargeswitch}(d) we show the NV$^{-}$ population as a function of the recharge pulse length, now at a (maximal) power of 1.1 $\mu$W. The population grows quickly, with a fitted time constant of 71(2) $\mu$s. However, beyond a maximum population of 88.6(6)\% at $t$ = 0.55 ms, the signal again decays. A possible explanation for this decay is spectral diffusion, caused by ionisation of nearby P1 centres by the yellow laser \cite{heremans2009generation}. Note that significantly higher recharging probabilities were achieved in Ref. \cite{baier2020orbital}, where the diamond substrate contained significantly fewer impurities. In Fig. \ref{fig:chargeswitch}(e), we focus on the first 500 $\mu$s of recharging, and probe the population in each of the NV$^{-}$ spin states. We observe that the recharging laser induces fast spin-pumping into the $m_{s}$ = 0 state, as expected due to the intersystem crossing. At $t$ = 500 $\mu$s, the fraction of NV$^{-}$ population in this state is 91.9(8)\%. Note that this spin-pumping effect mitigates dephasing of the nuclear spin in the time between recharging and resonant spin-reset of the NV$^{-}$ spin.

\FloatBarrier

\subsection{Simulated performance of a weakly-coupled data qubit for distributed quantum information processing}

In the main text, we present the simulated performance of two-qubit non-local operations and generation of four-qubit GHZ states between nuclear spin qubits hosted in disparate quantum network nodes. Here, we outline the models and methods used for these simulations. Further details can be found in the code implementation \cite{repo}.

\subsubsection{Simulation parameters}

In Tables \ref{tab:sim_parameters}-\ref{tab:purifiedSample}, we give the list of relevant parameters, which are described in the following sub-sections. Table \ref{tab:oldSample} describes gate durations which are specific to a natural abundance $^{13}$C device, alongside measured decoherence rates in such a device \cite{pompili2021realization}, whereas Table \ref{tab:purifiedSample} describes the numbers used and characterised in this work (0.01\% $^{13}$C abundance). 

\begin{table*}[h]
    \centering
    \begin{tabular}{l|l|l}
         \textbf{Parameter} & \textbf{Description} & \textbf{Values} \\\hline \hline
         \multicolumn{3}{c}{\textbf{General parameters}}\\ \hline\hline
        $\mathbf{p_g}$                 & Gate error probability & 0.01 \\ \hline
         $\mathbf{p_m}$                 & Measurement error probability & 0.01 \\\hline
         $\mathbf{p_n}$                 & Network error probability & 0.1 \\\hline $\mathbf{p_\textbf{RE}}$  & RE success probability & 0.0001 \\ \hline
         $\mathbf{T}_\textbf{meas}$     & Measurement duration   & $4.0\cdot 10^{-6}$ s \\\hline
         $\mathbf{T}_\textbf{RE}$      & RE duration             & $6.0\cdot 10^{-6}$ s \\\hline
         $\mathbf{T}^{e}_\mathbf{X}$ & Electron $X$ gate duration & $0.14 \cdot 10^{-6}$ s \\\hline
         $\mathbf{T}^{e}_\mathbf{Y}$ & Electron $Y$ gate duration & $0.14 \cdot 10^{-6}$ s \\\hline
         $\mathbf{T}^{e}_\mathbf{Z}$ & Electron $Z$ gate duration & $0.10 \cdot 10^{-6}$ s \\\hline
         $\mathbf{T}^{e}_\mathbf{H}$ & Electron $H$ gate duration &  $0.10 \cdot 10^{-6}$ s \\\hline
    \end{tabular}
    \caption{General system parameters as used for the simulations presented in Figs. 4(c,d) of the main text.}
    \label{tab:sim_parameters}
\end{table*}

\begin{table*}[h]
    \centering
    \begin{tabular}{l|l|l}
         \hline\hline
         \multicolumn{3}{c}{\textbf{$^{13}$C decoherence parameters (Ref. \cite{pompili2021realization})}}\\ \hline\hline
         $\mathbf{T1^{n}_\textbf{idle}}$ & $^{13}$C $T_{1}$ when idle & $300$ s \\\hline
         $\mathbf{T1^{n}_\textbf{RE}}$ & $^{13}$C $T_{1}$ during RE attempts & 0.03 s \\\hline
         $\mathbf{T2^{n}_\textbf{idle}}$ & $^{13}$C $T_{2}$ when idle & 10 s \\\hline
         $\mathbf{T2^{n}_\textbf{RE}}$ & $^{13}$C $T_{2}$ during RE attempts & 0.012 s \\\hline
         $\mathbf{T2^{e}_\textbf{idle}}$ & Electron $T_{2}$ when idle & 1.0 s \\\hline \hline
         \multicolumn{3}{c}{\textbf{1.1\% $^{13}$C gate durations}}\\ \hline\hline
         $\mathbf{T}^{C}_\mathbf{X}$ & $^{13}$C $X$ gate duration & $1.0 \cdot 10^{-3}$ s \\\hline
         $\mathbf{T}^{C}_\mathbf{Y}$ & $^{13}$C $Y$ gate duration & $1.0 \cdot 10^{-3}$ s  \\\hline
         $\mathbf{T}^{C}_\mathbf{Z}$ & $^{13}$C $Z$ gate duration & $0.5 \cdot 10^{-3}$ s  \\\hline
         $\mathbf{T}^{C}_\mathbf{H}$ & $^{13}$C $H$ gate duration & $0.5 \cdot 10^{-3}$ s \\\hline 
         $\mathbf{T}_\mathbf{CNOT}$ & CNOT gate duration & $0.5 \cdot 10^{-3}$ s \\\hline
         $\mathbf{T}_\mathbf{CZ}$ & CZ gate duration & $0.5 \cdot 10^{-3}$ s \\\hline
         $\mathbf{T}_\mathbf{SWAP}$ & SWAP gate duration & $1.5 \cdot 10^{-3}$ s
    \end{tabular}
    \caption{Decoherence timescales as measured in Ref. \cite{pompili2021realization}, and typical gate durations for such a device (1.1\% $^{13}$C) \cite{bradley2019ten}, as used for the simulations presented in Fig. 4(d) of the main text (labelled by $N_{\mathrm{1/e}}$ = 2$\cdot$10$^{3}$, the $T2^{n}_\mathrm{RE}$ decay).}
    \label{tab:oldSample}
\end{table*}

\begin{table*}[h]
    \centering
    \begin{tabular}{l|l|l}
         \hline\hline
         \multicolumn{3}{c}{\textbf{$^{13}$C decoherence parameters (this work)}}\\ \hline\hline
         $\mathbf{T1^{n}_\textbf{idle}}$ & $^{13}$C $T_{1}$ when idle & $300$ s \\\hline
         $\mathbf{T1^{n}_\textbf{RE}}$ & $^{13}$C $T_{1}$ during RE attempts & 1.2 s \\\hline
         $\mathbf{T2^{n}_\textbf{idle}}$ & $^{13}$C $T_{2}$ when idle & 10 s \\\hline
         $\mathbf{T2^{n}_\textbf{RE}}$ & $^{13}$C $T_{2}$ during RE attempts & 1.2 s \\\hline
         $\mathbf{T2^{e}_\textbf{idle}}$ & Electron $T_{2}$ when idle & 1.0 s \\\hline \hline
         \multicolumn{3}{c}{\textbf{0.01\% $^{13}$C gate durations}}\\ \hline\hline
         $\mathbf{T}^{C}_\mathbf{X}$ & $^{13}$C $X$ gate duration & $13 \cdot 10^{-3}$ s \\\hline
         $\mathbf{T}^{C}_\mathbf{Y}$ & $^{13}$C $Y$ gate duration & $13 \cdot 10^{-3}$ s \\\hline
         $\mathbf{T}^{C}_\mathbf{Z}$ & $^{13}$C $Z$ gate duration & $6.5 \cdot 10^{-3}$ s \\\hline
         $\mathbf{T}^{C}_\mathbf{H}$ & $^{13}$C $H$ gate duration & $6.5 \cdot 10^{-3}$ s \\\hline
         $\mathbf{T}_\mathbf{CNOT}$ & CNOT gate duration & $25 \cdot 10^{-3}$ s \\\hline
         $\mathbf{T}_\mathbf{CZ}$ & CZ gate duration & $25 \cdot 10^{-3}$ s \\\hline
         $\mathbf{T}_\mathbf{SWAP}$ & SWAP gate duration & $75 \cdot 10^{-3}$ s \\\hline
    \end{tabular}
    \caption{Decoherence timescales as measured in this work, and associated gate durations (0.01\% $^{13}$C device), as used for the simulations presented in Figs. 4(c) (see \emph{``Non-local CNOT operations''}) and 4(d) (labelled by $N_{\mathrm{1/e}}$ = 2$\cdot$10$^{5}$, the $T2^{n}_\mathrm{RE}$ decay).}
    \label{tab:purifiedSample}
\end{table*}

\subsubsection{Node topology}

We assume that the NV centre has a star topology, where two qubit gates are only possible between the electron spin qubit and another qubit. Moreover, CNOT gates can only be carried out controlled by the electron spin qubit. Direct read-out is only possible for the electron spin qubit: other qubits must be read-out by first mapping their state to the electron spin. The protocols considered here require a maximum of three $^{13}$C qubits in each node. 

\subsubsection{Decoherence}
Decoherence is modelled using two types of noise channels: the generalised amplitude damping channel and the phase damping channel \cite{nielsen2002quantum} which have characteristic timescales of ${T_1}$ and ${T_2}$ respectively. We assign different ${T_1}$ and ${T_2}$ values to the nuclear spins dependent on whether their associated electron spin is being used to generate inter-node entanglement at a given point in the sequence.

The set of Kraus operators for the generalised amplitude damping channel is given by
\begin{equation}
\begin{split}
    K_{AD1} &= \sqrt{\frac{1}{2}} \begin{bmatrix}
    1 & 0\\
    0 & \sqrt{1-\gamma_1}\\
    \end{bmatrix}, \\ 
    K_{AD2} &= \sqrt{\frac{1}{2}} \begin{bmatrix}
    0 & \sqrt{\gamma_1}\\
    0 & 0\\
    \end{bmatrix}, \\
    K_{AD3} &= \sqrt{\frac{1}{2}} \begin{bmatrix}
    \sqrt{1-\gamma_1} & 0\\
    0 & 1\\
    \end{bmatrix}, \\ 
    K_{AD4} &= \sqrt{\frac{1}{2}} \begin{bmatrix}
    0 & 0\\
    \sqrt{\gamma_1} & 0\\
    \end{bmatrix}.
\end{split}
\end{equation}
where $\gamma_1=1-\text{exp}(-t/T_1)$.

The set of Kraus operators for the phase damping channel is given by:
\begin{equation}
    K_{PD1} = \begin{bmatrix}
    1 & 0\\
    0 & \sqrt{1-\gamma_2}\\
    \end{bmatrix}, \ 
    K_{PD2} = \begin{bmatrix}
    0 & 0\\
    0 & \sqrt{\gamma_2}\\
    \end{bmatrix}.
\end{equation}
where $\gamma_2=1-\text{exp}(-t/\bar{T}_2)$.

In Tables \ref{tab:oldSample} and \ref{tab:purifiedSample}, the $T_{1}$ times correspond to the experimentally measured decoherence times of the eigenstates of the Pauli-Z matrix ($\ket{\uparrow}$,$\ket{\downarrow}$), and correspond directly to the $T_{1}$ times used for the generalised amplitude damping channel of the model. The $T_{2}$ times correspond to the experimentally measured decoherence times of the eigenstates of the Pauli-X and Pauli-Y matrices ($\ket{\mathrm{x}}$,$\ket{\mathrm{-x}}$,$\ket{\mathrm{y}}$,$\ket{\mathrm{-y}}$). To account for the dephasing induced by the generalised amplitude damping channel, modified $T_{2}$ values must be used in the phase damping channel, which are given by:
\begin{equation}		
\frac{1}{\bar{T}_2} = \frac{2}{T_2} - \frac{1}{T_1}		
\end{equation}

\subsubsection{Gates}

We model single-qubit gates as noiseless. We model noisy two-qubit gates using a depolarizing noise model: 
\begin{equation}
    N_{2Q}(\rho, p_g) = (1-p_g)\rho + \frac{p_g}{15}\sum_{\sigma_i,\sigma_j} (\sigma_i \otimes \sigma_j) \rho (\sigma_i \otimes \sigma_j)^\dagger.
\end{equation}
where $\sigma_i,\sigma_j\in \{I, X, Y, Z\}$ and the sum is over all combinations of the Pauli operators $\sigma_i,\sigma_j$ except for $\sigma_i=\sigma_j=I$.

For simplicity of compilation, we consider a native gate-set consisting of the Pauli gates, the Hadamard gate, the CNOT gate and the CZ gate. SWAP gates are carried out using three CNOT gates. All single- and two-qubit gates have associated durations: after each operation, the appropriate ${T_1}$ and ${T_2}$ decoherence is applied to all idle qubits.

\subsubsection{NV-NV entanglement}
We model the creation of remote-entanglement (RE) links (i.e. NV-NV Bell pairs) following the single-click protocol \cite{cabrillo1999creation}. The protocol produces states that we assume are of the form
\begin{equation}
    \rho_{NV,NV} = (1-p_n)\ket{\psi}\bra{\psi} + p_n \ket{11}\bra{11}.
    \label{eq:LDE}
\end{equation}
with $\ket{\psi}=(\ket{01}+\ket{10})/\sqrt2$ and $p_n$ a parameter that we call the network error, arising from events in which both NVs emitted a photon. That is, we assume the intrinsic infidelity associated with the single-photon scheme dominates over other error sources such as imperfect visibility \cite{humphreys2018deterministic}. The single-click protocol succeeds with probability $p_\mathrm{RE}$.

In simulation, each time a RE link must be generated, Monte-Carlo methods are used to determine the number of entangling attempts $n_{\mathrm{RE}}$ before success. The decoherence channels are then applied to all idle qubits with an associated duration of $t_{\mathrm{ent}}$ = $n_{\mathrm{RE}} T_{RE}$, where $T_{RE}$ is the duration of each entangling primitive.

The results presented in Fig. 4(c) of the main text are averaged over 500,000 repetitions, whereas those in Fig. 4(d) use 100,000 repetitions, aside from the Expedient protocol for $N_{\mathrm{1/e}}$ = 2$\cdot$10$^{3}$, for which only 10,000 repetitions are used due to the low protocol success rate.

\subsubsection{Measurement noise}
We restrict measurements to the $Z$-basis. To measure the qubits in the $X$-basis, we apply a Hadamard gate before the measurement. The gate duration time of this additional operation is included in the simulations.

We model (single qubit) measurement errors by projecting onto the opposite eigenstate of the measured operator. Measurement errors are characterised by the probability $p_m$ of an incorrect projection. 

\subsubsection{Dynamical decoupling}

To mitigate dephasing of both communication and data qubits due to quasi-static noise sources (for example, due to spin-bath interactions or magnetic field fluctuations), a dynamical decoupling scheme is accounted for in the sequencing of all operations. We model this process by only applying operations at intervals of a basic echo sequence on the data qubits: $\tau - \pi - \tau$. Here, $\pi$ denotes a $\pi$-pulse being applied to the system, and $\tau$ is the inter-pulse delay time. Operations are only allowed in between the echo sequences of each node; consecutive operations in the same node are allowed, after which the decoupling sequence is restarted. We set $\tau$ equal to the duration of $2500$ entangling attempts for the purified sample and a $\tau$ equal to the duration of $250$ entangling attempts for the natural abundance sample. The $\pi$-pulse duration is set to $13\cdot 10^{-3}$ seconds for the purified sample and $1\cdot 10^{-3}$ seconds for the natural abundance sample. Due to the faster gate operations associated with the communication qubits, we assume that refocusing points of their own decoupling sequences can be simply matched to those of the data qubits. 

\subsubsection{Non-local CNOT operations}

In Fig. 4(c) of the main text, we present the average fidelity of a non-local CNOT operation. Here, we use the parameters given in Tables \ref{tab:sim_parameters} and \ref{tab:purifiedSample}, but sweep the values of $T1^{n}_{RE}=T2^{n}_{RE}=N_{\mathrm{1/e}} \cdot T_{RE}$. The average fidelity of the operation is calculated by determining the entanglement fidelity following the relation given in Ref. \cite{nielsen2002simple}. We do not show the results when using the faster gate and measurement durations (Table \ref{tab:oldSample}), because a negligible difference is observed. For example, for $N_{\mathrm{1/e}}$ = 83,333, we find an average fidelity of 0.7899, compared with 0.7890 when using the slower durations.

\subsubsection{GHZ-creation protocols}

\begin{figure*}[h]
    \makebox[\textwidth][c]{\includegraphics[width=1\textwidth]{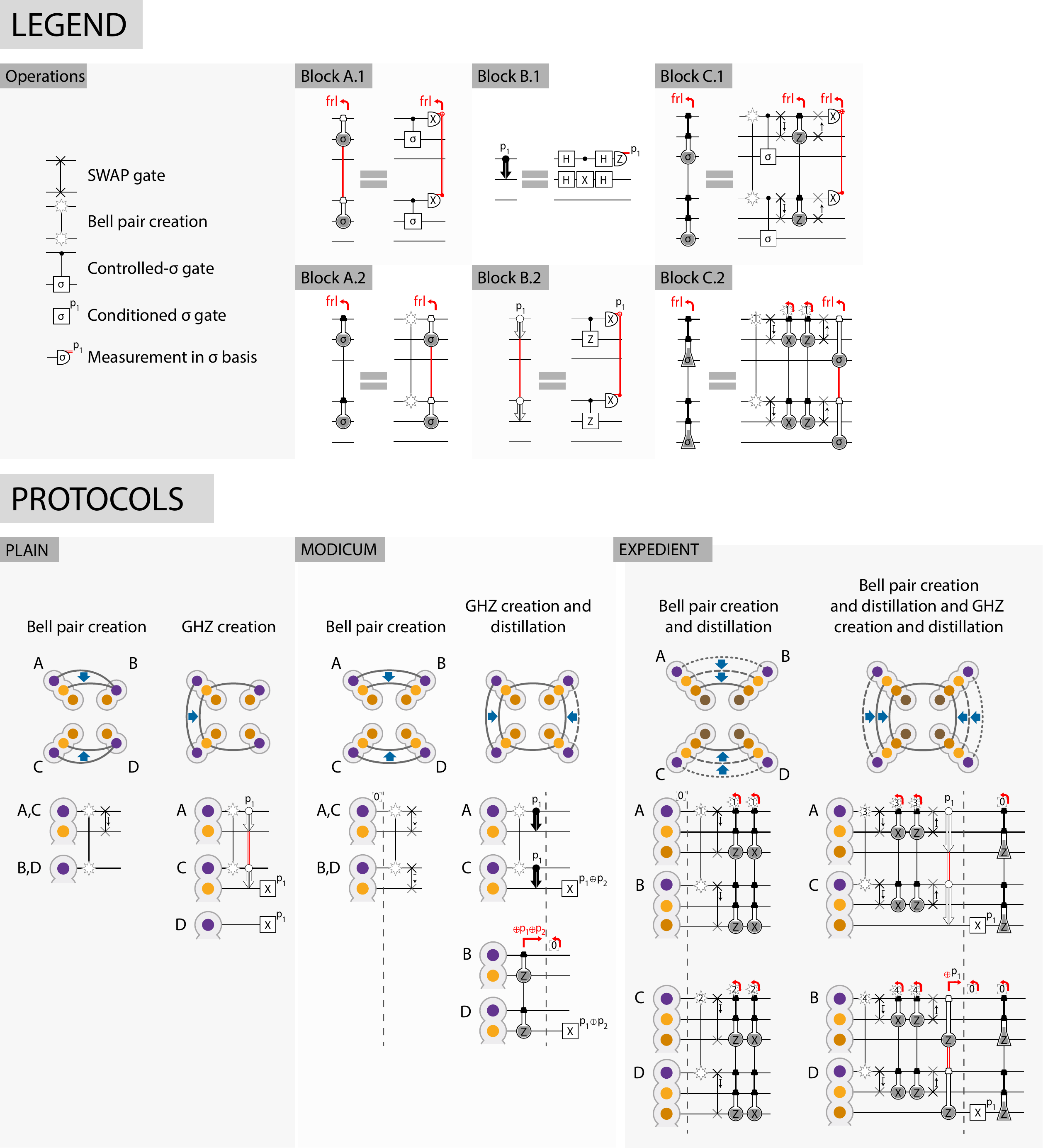}}
    \caption{Circuit diagrams of the three GHZ creation protocols used in the simulations. Bell states are generated in the form of Eq. \ref{eq:LDE}, followed by an X-gate on their second qubit to turn them into noisy versions of the $(\ket{00}+\ket{11})/\sqrt{2}$ state. `Plain' uses three Bell pairs to create a 4-qubit GHZ state. `Modicum' uses a fourth Bell pair to increase the GHZ fidelity with a distillation step. `Expedient' \cite{nickerson2013topological} uses 22 Bell pairs and has many (intermediate) distillation steps. Details of those steps are given in the text. The schematics above the circuits show four network nodes (A,B,C,D) comprising an electron spin (purple) and up to three nuclear spins (orange tones).}
    \label{fig:GHZcircuits}
\end{figure*}

In this section, we discuss the three GHZ creation protocols used in Fig. 4(d) of the main text. The circuit diagrams of the protocols are depicted in Fig. \ref{fig:GHZcircuits}. In each node, the first (purple) qubit indicates the electron qubit of the NV centre (i.e., the communication qubit), with the other qubits indicating $^{13}$C spins (`memory' and `data' qubits). The memory qubits are used to store intermediate states required to create the GHZ state. When the GHZ state is created, it can be consumed to non-locally perform a weight-4 Pauli measurement on qubits of a distributed error-correcting code, e.g., the surface code \cite{nickerson2013topological,nickerson2014freely}. In the diagrams of Fig. \ref{fig:GHZcircuits}, the inner-most $^{13}$C is the data qubit of the code (i.e., it would be a constituent of an encoded logical qubit).

The Plain protocol is the most simple protocol to create a 4-qubit GHZ state out of Bell pairs: it only requires three pairs. The local operations of this protocol `fuse' the three pairs into the GHZ state. Fusion takes place with Block B.2 from the legend: this operation fuses two entangled states that do not overlap in any of the nodes with the aid of a Bell pair that is consumed in the process. Feed-forward operations based on this Bell-state measurement lead to protocol success with unit probability.

As mentioned above, RE links can only be created between the electron qubits. Hence, the states of the electron qubits must be swapped with those of data qubits (`stored') at some point in all of the protocols. In the Plain protocol, this occurs after the first pair of RE links are created: the electron qubits of nodes $A,C$ are stored such that it is possible to generate a third RE link between nodes $A$ and $C$.

The Modicum protocol uses one additional Bell pair compared to Plain. This additional pair is used to distill the 4-qubit GHZ state. This distillation step is probabilistic; if successful it increases the state fidelity. However, if it fails, the GHZ state is discarded and the protocol restarts from scratch. Fusion of the Bell pairs takes place in Modicum with Block B.1 (both in nodes $A$ and $C$ in Fig. \ref{fig:GHZcircuits}), which describes fusion of two entangled states that overlap in one network node. Distillation takes place with Block A.2 in nodes $B$ and $D$, which essentially measures a stabilizer of the desired pure entangled state. In particular, Block A.2 is known in the literature as `single selection' \cite{fujii2009entanglement}. The operation is successful if the measurement results are even: this corresponds to the +1 outcome of the stabilizer measurement. In case of an odd measurement combination, the distillation operation fails, and the protocol returns to the indicated `failure reset level' (indicated with `frl' in the legend). The Modicum protocol is identified with the dynamic programming algorithm presented in \cite{de2020protocols}, and is similar to the protocol that creates one of the two GHZ states of the `Basic' protocol in \cite{nickerson2014freely}.

The Expedient protocol \cite{nickerson2013topological} uses 22 Bell pairs to create the GHZ state. Compared to the Plain protocol, the additional 19 Bell pairs are used for distillation purposes. This can be either for distillation of Bell pairs before they are fused to form the GHZ state, or for distillation of the GHZ state after it is created. Distillation takes place with Block A.2, but also with Block C.1 and Block C.2. Block C.1 can be understood as an operation that performs two distillation steps by consuming two ancillary Bell pairs. This block is known in the literature as the `double selection' distillation protocol \cite{fujii2009entanglement}. Block C.2 essentially performs three distillation steps by consuming three Bell pairs. Depending on where in the circuit the distillation steps occur, failure of a step means the entire protocol has to start from scratch, or only a specific branch of the protocol has to be carried out again. Finally, similar to the Plain protocol, this protocol uses Block B.2 to fuse three Bell pairs into a GHZ state. This takes place between distillation blocks A.2 and C.2.

\clearpage
\bibliography{Qmem}